\documentclass{article}

\newlength{\abstractwidth}
\usepackage{cancel}
\usepackage{hyperref}
\usepackage{color}
\usepackage{graphicx}
\usepackage{cite}

\usepackage{tikz, pgf}
\usepackage{tkz-fct}
\usepackage{pgfplots}
\usetikzlibrary{shapes.misc}
\usetikzlibrary{shapes,snakes}
\usetikzlibrary{decorations.pathmorphing}	
\usetikzlibrary{decorations.markings}

\usetikzlibrary{arrows.meta}

\usepackage{amsmath}
\usepackage{amssymb}
\usepackage{latexsym}
\usepackage{physics}
 \usepackage{setspace}
 
 \usepackage{caption}
\usepackage{subcaption}

\numberwithin{equation}{section}

\renewcommand{\thefootnote}{\fnsymbol{footnote}}
\renewcommand{\thanks}[1]{\footnote{#1}}
\newcommand{\starttext}{
\setcounter{footnote}{0}
\renewcommand{\thefootnote}{\arabic{footnote}}}
\newcommand{\bea}{\begin{eqnarray}}
\newcommand{\eea}{\end{eqnarray}}
\newcommand{\be}{\begin{eqnarray}}
\newcommand{\ee}{\end{eqnarray}}

\def\ie{\begin{equation}\begin{aligned}}
\def\fe{\end{aligned}\end{equation}}

\def\ie{\begin{equation}\begin{aligned}}
\def\fe{\end{aligned}\end{equation}}

\def\cC{{\cal C}}

\def\Tr{{\rm Tr}}

\begin{document}

\starttext

\setcounter{footnote}{0}

\begin{flushright}
{\small QMUL-PH-23-02}
\end{flushright}

\vskip 0.3in

\begin{center}

\centerline{\large \bf  Laplace-difference equation for integrated correlators  }
\vskip 0.03in
\centerline{\large \bf of operators  with general charges in $\mathcal{N}=4$ SYM} 

\vskip 0.2in

{Augustus Brown, Congkao Wen, and Haitian Xie} 
   
\vskip 0.15in

{\small   Centre for Theoretical Physics, Department of Physics and Astronomy,  }\\ 
{\small Queen Mary University of London,  London, E1 4NS, UK}

\vskip 0.15in

{\tt \small a.a.x.brown@qmul.ac.uk,   c.wen@qmul.ac.uk, h.xie@se21.qmul.ac.uk}

\vskip 0.5in

\begin{abstract}
 
 \vskip 0.1in

We consider the integrated correlators associated with four-point correlation functions $\langle \mathcal{O}_2\mathcal{O}_2\mathcal{O}^{(i)}_p \mathcal{O}^{(j)}_p  \rangle$ in four-dimensional $\mathcal{N}=4$ supersymmetric Yang-Mills theory (SYM) with $SU(N)$ gauge group, where $\mathcal{O}^{(i)}_p$ is a superconformal primary with charge (or dimension) $p$ and the superscript $i$ represents possible degeneracy.  These integrated correlators are defined by integrating out spacetime dependence with a certain integration measure, and they can be computed via supersymmetric localisation. They are modular functions of complexified Yang-Mills coupling $\tau$. We show that the localisation computation is systematised by appropriately reorganising the operators. After this reorganisation of the operators, we prove that all the integrated correlators for any $N$, with some crucial normalisation factor, satisfy a universal Laplace-difference equation (with the laplacian defined on the $\tau$-plane) that relates integrated correlators of operators with different charges. This Laplace-difference equation is a recursion relation that completely determines all the integrated correlators, once the initial conditions are given. 

\end{abstract}                                            
\end{center}

\baselineskip=15pt
\setcounter{footnote}{0}

\newpage

\setcounter{page}{1}
\tableofcontents

\newpage

\section{Introduction}

It has been shown in \cite{Binder:2019jwn,Chester:2020dja} (see also \cite{Chester:2019pvm, Chester:2019jas, Chester:2020vyz}) that the four-point correlation function of superconformal primary operators of the stress tensor multiplet  in $\mathcal{N}=4$ supersymmetric Yang-Mills (SYM) theory can be computed exactly when we integrate out the spacetime dependence with a certain choice of the measure. Such physical observables are called integrated correlators and they are related to the partition function of $\mathcal{N}=2^{\star}$ SYM on $S^4$, which can be computed using supersymmetric localisation \cite{Pestun:2007rz}. There are two known choices of the integration measure \cite{Binder:2019jwn,Chester:2020dja}. For one of these measures, the integrated correlator is given by 
\ie \label{eq:2222}
\cC_{G_N}(\tau, \bar{\tau}) = {1\over 4} \Delta_\tau \partial_{m}^2 \log \mathcal{Z}_{G_N} (\tau, \bar{\tau}; m) \big{|}_{m=0} \, ,
\fe  
where $\tau=\theta/(2\pi) + i \, 4\pi/g^2_{_{YM}}  = \tau_1 + i\, \tau_2$ is the complexified Yang-Mills coupling, $\Delta_\tau = 4 \tau_2^2 \partial_{\tau}\partial_{\bar \tau}$ is the hyperbolic laplacian, and $\mathcal{Z}_{G_N}$ is the partition function of $\mathcal{N}=2^{\star}$ SYM with gauge group $G_N$ on $S^4$, which can be viewed as $\mathcal{N}=4$ SYM deformed by some mass terms (denoted as $m$ in the above formula) in the hypermultiplet. Due to supersymmetric localisation, the partition function $\mathcal{Z}_{G_N}$  can be expressed as an $N$-dimensional matrix-model integral over Coulomb branch parameters with an integrand that contains both perturbative and non-perturbative instanton contributions \cite{Pestun:2007rz, Nekrasov:2002qd}. It was found in \cite{Dorigoni:2021bvj,Dorigoni:2021guq, Dorigoni:2022zcr} that $\cC_{G_N}(\tau, \bar{\tau})$ obeys a set of recursion relations, which were called Laplace-difference equations, that relate the integrated correlators of different gauge groups\footnote{See \cite{Dorigoni:2022iem} for a recent SAGEX review on this topic.}. For example, in the case of $SU(N)$ that we will consider in this paper, the Laplace-difference equation takes the following form 
\ie \label{eq:N-LG}
 \Delta_{\tau}   &\,  \cC_{N}(\tau, \bar \tau) = N(N-1)\, \cC_{N+1}(\tau, \bar \tau) -2(N^2-1)\, \cC_{N}(\tau, \bar \tau) +N(N+1)\, \cC_{N-1}(\tau, \bar \tau) \, ,
\fe
 where to simplify the notation we have denoted $\cC_{SU(N)}(\tau, \bar \tau)$ as $\cC_{N}(\tau, \bar \tau)$. 

Following the earlier proposal \cite{Binder:2019jwn}, the supersymmetric localisation computation has been recently applied to more general correlators of the form $\langle \mathcal{O}_2 \mathcal{O}_2 \mathcal{O}^{(i)}_{p} \mathcal{O}^{(j)}_{p}  \rangle$ \cite{Paul:2022piq},\footnote{Note the four-point correlator $\langle \mathcal{O}_2\, \mathcal{O}_2\, \mathcal{O}^{(i)}_p\, \mathcal{O}^{(j)}_q  \rangle$ is non-trivial in $\mathcal{N}=4$ SYM only if $p=q$, see e.g. \cite{Rayson:2008uje, Aprile:2020uxk, DHoker:2000xhf}.}  where $\mathcal{O}^{(i)}_{p}$ is a superconformal primary of dimension $p$ that is charged under $SU(4)$ R-symmetry (so we will call $p$ charge or dimension interchangeably).\footnote{See also \cite{Fiol:2023cml} for recent related work on these integrated correlators in the planar limit.} In general, there are multiple operators which have the same dimension $p$ (when $p>3$), and the superscript $i$ is to label the degeneracy. We will denote these more general integrated correlators as $\cC^{(i, j)}_{N, p}(\tau; \bar{\tau})$, and they are modular functions of $(\tau, \bar{\tau})$ due to Montonen-Olive duality of $\mathcal{N}=4$ SYM with $SU(N)$ gauge group \cite{Montonen:1977sn}. When $p=2$, it reduces to the integrated correlator $\cC_{N}(\tau, \bar \tau)$ that appeared in \eqref{eq:2222} and \eqref{eq:N-LG}. It was argued in \cite{Binder:2019jwn, Paul:2022piq} that ${\cC}^{(i,j)}_{N,p}(\tau; \bar{\tau})$ for $p>2$ can be related to the partition function of $\mathcal{N}=2^{\star}$ SYM on $S^4$ deformed by the higher-dimensional operators $\mathcal{O}^{(i)}_p$, which again can be computed via supersymmetric localisation. Certain examples of integrated correlators for particular values of $p$ as well as a particular class of operators (called maximal-trace operators) were explicitly studied in \cite{Paul:2022piq}.  Remarkably, as demonstrated in \cite{Paul:2022piq}, recurrence relations were found for certain examples of integrated correlators, which relate integrated correlators of different ranks of the gauge group as well as different charges, generalising the Laplace-difference equation given in \eqref{eq:N-LG}.

In this paper, we will prove that the integrated correlators ${\cC}^{(i,j)}_{N,p}(\tau; \bar{\tau})$ for any $p$ and $N$ obey a universal Laplace-difference equation that relates the integrated correlators of operators with different charges. The localisation computation is performed on $S^4$, where there is mixing for operators with different dimensions due to the dimensionful radius of $S^4$. The mixing is resolved through the Gram-Schmidt procedure \cite{Gerchkovitz:2016gxx, Binder:2019jwn}. To systematise the Gram-Schmidt procedure and simplify the localisation computation, it is important to reorganise the operators, as we describe in section \ref{sec:re-org}.  After this crucial reorganisation, the four-point functions now take the following form, 
\ie \label{eq:cla}
\langle \mathcal{O}_2\, \mathcal{O}_2 \, \mathcal{O}^{(i)}_{p|M} \, \mathcal{O}^{(i')}_{p'|M'} \rangle \, , \qquad {\rm with} \qquad \mathcal{O}^{(i)}_{p|M}:= (\mathcal{O}_2)^p \mathcal{O}^{(i)}_{0|M}\,,
\fe
where we have reorganised the higher-dimensional operators, which are now denoted as $\mathcal{O}^{(i)}_{p|M}$. Here $M=0, 3, 4, \ldots$, and superscript $i$ denotes possible degeneracy for a given $M$, which happens when $M\geq 6$. We will simply omit the index $i$ for $M<6$. When $M=0$, $\mathcal{O}_{0|0}=\mathbb{I}$ is the identity operator, and in general $\mathcal{O}^{(i)}_{0|M}$ is a dimension-$M$ operator that is defined recursively in \eqref{eq:defOm}. Therefore $\mathcal{O}^{(i)}_{p|M}$ has dimension $2p+M$, and for the correlators to be non-trivial, we must have $2p+M=2p'+M'$ (so $p'$ is not an independent variable).  We will denote the integrated correlators associated with the four-point functions given in \eqref{eq:cla} as $ \cC_{N,p}^{(M, M'|i, i')}(\tau, \bar \tau)$. It will be shown that it is crucial  to normalise the integrated correlators appropriately, and the integrated correlators including this important normalisation factor will be denoted as ${\widehat \cC_{N,p}}^{(M, M'|i, i')}(\tau, \bar \tau)$. 

We will then prove that remarkably ${\widehat \cC_{N,p}}^{(M, M'|i, i')}(\tau, \bar \tau)$ obeys a universal Laplace-difference equation that relates the integrated correlators with different charges $p$,\footnote{To be precise the total charge of $\mathcal{O}^{(i)}_{p|M}$ is $2p+M$. We will mostly be interested in the situation with fixed $M$ and different $p$'s, so we will loosely call $p$ the charge.} which takes the following form, 
\ie \label{eq:gen-LDint}
\Delta_{\tau}\, {\widehat \cC_{N,p}}^{(M, M'|i, i')} & (\tau, \bar \tau)  =\, \left(p+1+ \delta  \right)\left(p+ a+1 \right) {\widehat \cC_{N,p+1}}^{(M, M'|i,i')}(\tau, \bar \tau) + p\left(p+ a +\delta \right) {\widehat \cC_{N,p-1}}^{(M, M'|i,i')}(\tau, \bar \tau) \cr
& - \left[ 2p\left( p+ a\right) + (2p+a+1)(\delta+1) \right] {\widehat \cC_{N,p}}^{(M, M'|i,i')}(\tau, \bar \tau) -4 \, \delta_{M,M'} \delta_{i, i'}\, \cC_{N}(\tau, \bar \tau) \, ,
\fe
where 
\ie 
a ={N^2+M+M'-3\over 2}  \,, \qquad  \qquad \delta = {M - M' \over 2} \, ,
\fe
and the ``source term" $\cC_{N}(\tau, \bar \tau)$ is the integrated correlator that appeared in \eqref{eq:2222}, whose expression is known exactly, and  obeys its own Laplace-difference equation as given in \eqref{eq:N-LG} \cite{Dorigoni:2021bvj,Dorigoni:2021guq}.\footnote{Note $\cC_{N}(\tau, \bar \tau)$ is the integrated correlator associated with four-point function $\langle \mathcal{O}_2 \mathcal{O}_2 \mathcal{O}_2 \mathcal{O}_2 \rangle$, so in our notation it is equivalent to $\cC_{N,1}^{(0,0)}(\tau, \bar \tau)$. Namely, this corresponds to $M=M'=0$ and $p=1$, for which there is no degeneracy. To unify the notation, we will denote $\cC_{N}(\tau, \bar \tau)$ as $\cC_{N,1}^{(0,0)}(\tau, \bar \tau)$, where we have omitted the indices $i, i'$. Furthermore, as will become clear later, we have $\cC_{N,1}^{(0,0)}(\tau, \bar \tau) = {N^2-1 \over 8} \widehat{\cC}_{N,1}^{(0,0)}(\tau, \bar \tau)$.} . The Laplace-difference equation \eqref{eq:gen-LDint} is a recursion relation that completely determines ${\widehat \cC_{N,p}}^{(M, M'|i, i')}  (\tau, \bar \tau)$ for any $p$ in terms of the initial condition ${\widehat \cC_{N,0}}^{(M, M'|i, i')}  (\tau, \bar \tau)$.

The rest of the paper is organised as follows. In section \ref{sec:rev}, we will introduce the integrated correlators involving operators with general charges and their computation using supersymmetric localisation.  The localisation computation is performed on $S^4$, which leads to mixing of operators with different dimensions. Such mixing is resolved by a standard Gram-Schmidt procedure. We will reorganise the operators into towers as well as subtowers, which are associated with $M$ and $i$ in \eqref{eq:cla} respectively. Crucially, the operators in different (sub)towers have a vanishing two-point function on $S^4$, which implies the Gram-Schmidt procedure only needs to apply to the set of operators in a given (sub)tower. However, precisely because of the vanishing of these two-point functions, a naive definition of integrated correlators from localisation following \cite{Aprile:2020uxk} would lead to $0/0$ for the integrated correlators involving operators in different (sub)towers. We will carefully treat this issue and appropriately normalise the integrated correlators, so that $0/0$ is avoided and symmetry properties of the integrated correlators are made manifest.  In section \ref{sec:single}, we will prove that the integrated correlators with the normalisation factor defined in section \ref{sec:rev}  satisfy  the universal Laplace-difference equation \eqref{eq:gen-LDint}. We will find that the precise form of the partition function of $\mathcal{N}=2^*$ SYM that enters in the localisation computation of the integrated correlators is in fact not important for the integrated correlators  to obey the Laplace-difference equation.  The validity of the equation relies only on the reorganisation of operators and the Gram-Schmidt procedure, as well as some very basic properties of  the partition function such as its power behaviour as a function of $\tau_2$. In section \ref{sec:local}, we will give some details of the supersymmetric localisation computation and provide explicit perturbative results of certain  integrated correlators.  In particular, we will find that interestingly the perturbative contributions to integrated correlators involving operators in different (sub)towers start at higher loops. We will mainly focus on the perturbative contributions, but we will also comment on the $SL(2, \mathbb{Z})$ completion of the integrated correlators  and their lattice-sum representation.  We will conclude and comment on future directions in section \ref{sec:conclusion}.  The paper also includes an appendix \ref{app:Nrec}, in which we will provide exact expressions for the initial conditions (i.e. the integrated correlators with $p=0$) of the Laplace-difference equation for the integrated correlators that are considered in section \ref{sec:local}.

\vspace{0.3cm}

{\bf Note added}: While we were finalising this paper we learned about \cite{Erictoapp} where the large-charge properties of integrated correlators with maximal-trace operators were analysed.

\section{Integrated correlators of operators with general charges} \label{sec:rev}

In this section, we will briefly review the definition of the integrated correlators for the four-point correlation functions $\langle \mathcal{O}_2\mathcal{O}_2\mathcal{O}^{(i)}_p\mathcal{O}^{(j)}_p \rangle$, and their computation using supersymmetric localisation. Because the localisation computation is on $S^4$,  there is a mixing for operators with different dimensions, which is resolved through a Gram-Schmidt procedure \cite{Gerchkovitz:2016gxx, Binder:2019jwn}. To systematise the Gram-Schmidt procedure and simplify the localisation computation, it is crucial to reorganise the operators appropriately \cite{Gerchkovitz:2016gxx}. The operators will be organised into different towers, and each tower may further contain subtowers. We will perform the Gram-Schmidt procedure and determine the mixing coefficients that are associated with these reorganised operators. We will then introduce the normalisation factor for the integrated correlators, which will play an extremely important role in obtaining a universal Laplace-difference equation for the integrated correlators.

\subsection{Integrated correlators from localisation}
\label{sec:rev-local}

We are interested in correlation functions of four half-BPS superconformal primary operators in $\mathcal{N}=4$ SYM. These operators will be denoted as $\mathcal{O}^{(i)}_p$, where $p$ refers to the scaling dimension that is protected by supersymmetry. The superscript $i$ denotes the degeneracy, which happens when $p \geq 4$. Often, the superscript $i$ is omitted if there is no  degeneracy.  The operator $\mathcal{O}^{(i)}_p$ transforms in the $[0,p,0]$ representation of the $SU(4)$ R-symmetry, therefore we also may refer to $p$ as charge. Explicitly, in terms of fundamental scalar fields $\Phi^I$ (here $I=1,\cdots, 6$ is the $SO(6)$ R-symmetry index), the operators have the following forms: for the single-trace operators, we have\footnote{Here for the normalisation of the operators, we follow the convention of \cite{Paul:2022piq}.}
\ie  \label{eq:sin-tr}
T_p(x, Y) = {1\over p} Y_{I_1} \cdots Y_{I_p} \Tr\left( \Phi^{I_1}(x) \cdots \Phi^{I_p}(x) \right) \, ,
\fe
where $Y_{I}$ is a null $SO(6)$ vector;  the multi-trace operators are given by
\ie  \label{eq:mul-tr}
T_{p_1, \cdots, p_n}(x, Y) = {p_1 \cdots p_n\over p} T_{p_1}(x, Y) \cdots T_{p_n}(x, Y) \, ,
\fe
where $p=\sum_{i=1}^n p_i$. It is clear that for a given dimension $p \geq 4$, there are in general more than one distinguishable operator. For example, $\mathcal{O}_2$, which appears in the definition of the integrated correlators, is simply $T_2$, and at dimension four, one may choose  $\mathcal{O}^{(1)}_4=T_4$ and  $\mathcal{O}^{(2)}_4=T_{2,2}$. 

We now consider the correlation functions of these operators. Two- and three-point correlation functions are protected by supersymmetry \cite{Baggio:2012rr}, and the first non-trivial case is the four-point correlators. We will consider a special class of correlators of the form
$\langle \mathcal{O}_2\, \mathcal{O}_2\, \mathcal{O}^{(i)}_p\, \mathcal{O}^{(j)}_p  \rangle$.
Because of the constraints from superconformal
symmetry, the correlators can be expressed as \cite{Eden:2000bk, Nirschl:2004pa}
\ie \label{eq:22pp}
    \langle \mathcal{O}_2(x_1, Y_1)\mathcal{O}_2(x_2, Y_2)\mathcal{O}^{(i)}_p(x_3, Y_3)\mathcal{O}^{(j)}_p(x_4, Y_4) \rangle = \mathcal{G}_{\rm free}^{(i,j)}(x, Y) + \mathcal{I}_4 (x, Y)\mathcal{H}_{N, p}^{(i,j)} (U, V; \tau, \bar \tau) \, ,
\fe
where we have separated the result of the correlator into a free part, $\mathcal{G}_{\rm free}^{(i,j)}$, that is independent of the coupling and can be computed simply by free-field Wick contractions, and a part that non-trivially depends on the dynamics of the theory. For the dynamic part, one can further factor out the R-symmetry dependence, denoted $\mathcal{I}_4 (x, Y)$ (the expression can be found in \cite{Binder:2019jwn, Paul:2022piq}), which is completely fixed by the symmetry. The function $\mathcal{H}_{N, p}^{(i,j)} (U, V; \tau, \bar \tau)$ is our main focus, which  is a modular function of the complexified Yang-Mills coupling $\tau$ and the cross ratios, 
\ie
U= {x_{12}^2 x_{34}^2 \over x_{13}^2 x_{24}^2} \, , \quad \qquad V= {x_{14}^2 x_{23}^2 \over x_{13}^2 x_{24}^2} \, .
\fe
It also depends on the rank of gauge group $N$, the dimension of the operator $p$, and the degeneracy indices $(i,j)$. 

As shown in \cite{Binder:2019jwn} (see also \cite{Paul:2022piq} for further clarification, including the non-perturbative instanton effects), this particular type of correlator can be computed by supersymmetric localisation, once the space-time dependence has been integrated out with a certain measure. More concretely, we define the integrated correlators as \cite{Binder:2019jwn} 
\ie \label{eq:measure}
\cC_{N, p}^{(i,j)}(\tau, \bar \tau) = -{2\over \pi } \int_0^{\infty} dr \int^{\pi}_0 d \theta {r^3 \sin^2 \theta \over U^2} \mathcal{H}_{N, p}^{(i,j)} (U, V; \tau, \bar \tau) \, , 
\fe
with $U=1-2r \cos(\theta) +r^2$ and $V=r^2$, and $\cC_{N, p}^{(i,j)}(\tau, \bar \tau)$ can be related to the partition function $\mathcal{Z}_N(\tau, \tau_A'; m)$ of $\mathcal{N}=2^{\star}$ SYM on $S^4$ (now deformed by higher-dimensional operators $\mathcal{O}^{(i)}_p$) in the following manner. It is convenient to factorise  $\cC_{N, p}^{(i,j)}(\tau, \bar \tau)$ into the following form \cite{Paul:2022piq}, 
\ie \label{eq:CRCh}
\cC_{N, p}^{(i,j)}(\tau, \bar \tau) = {R_{N,p}^{(i,j)} \over 4}\, \widehat{\cC}_{N, p}^{(i,j)}(\tau, \bar \tau)\, , 
\fe
where the prefactor $R_{N,p}^{(i,j)}$ is independent of $\tau$, and is determined by the two-point function via
\ie \label{eq:RR}
\langle \mathcal{O}^{(i)}_p(x_1, Y_1) \, \mathcal{O}^{(j)}_p(x_2, Y_2) \rangle = R_{N,p}^{(i,j)}  \left( \frac{Y_1 \cdot Y_2}{ x_{12}^2} \right)^p\,,
\fe
and $\widehat{\cC}_{N, p}^{(i,j)}(\tau, \bar \tau)$, which is a highly non-trivial function of $(\tau, \bar \tau)$, can be obtained from the partition function $\mathcal{Z}_N(\tau, \tau_A'; m)$, 
\ie  \label{eq:C-hat}
\widehat{\cC}_{N, p}^{(i,j)}(\tau, \bar \tau) =  \frac{v_p^{i, \mu} \bar{v}_p^{j, \nu} \partial_{\tau'_{\mu}}\partial_{\bar{\tau}'_{\nu}} \partial_m^2\log \mathcal{Z}_N(\tau, \tau_A'; m) \big{|}_{m=\tau_A'=0} }{v_p^{i, \mu} \bar{v}_p^{j, \nu} \partial_{\tau'_{\mu}}\partial_{\bar{\tau}'_{\nu}} \log \mathcal{Z}_N(\tau, \tau_A'; 0) \big{|}_{\tau_A'=0} }\, .
\fe
Note in the above expression the repeated indices $\mu, \nu$ are summed. The vectors $v_p^{i, \mu}\, , \bar{v}_p^{j, \nu}$ are the so-called mixing coefficients, which are named as such due to the fact that on $S^4$,  operators with different dimensions can mix because of the additional dimensionful parameter, i.e. the radius of $S^4$ \cite{Gerchkovitz:2016gxx}. We will discuss the mixing coefficients in more detail later.

The connection between the integrated correlators and the partition function of $\mathcal{N}=2^*$ SYM can be roughly understood as follows. Two derivatives in mass $m$ bring down two $\mathcal{O}_2$'s, and the derivatives  $v_p^{i, \mu} \partial_{\tau'_{\mu}}$ and $ \bar{v}_p^{j, \nu} \partial_{\bar{\tau}'_{\nu}}$ lead to the associated dimension-$p$ operators $\mathcal{O}^{(i)}_p$ and $\mathcal{O}^{(j)}_p$. Since all the operators are brought down from the action, they are integrated over spacetime. The partition function $\mathcal{Z}_N(\tau, \tau_A'; m)$ is computable by supersymmetric localisation, and can be expressed as \cite{Pestun:2007rz}
\ie \label{eq:partition}
\mathcal{Z}_N(\tau, \tau_A'; m) =& \int d^{N-1}a  \left \vert \exp\left( i \pi \tau \sum_{i=1}^N a_i^2 +i \sum_{p>2} \pi^{p/2} \tau'_p \sum_{i=1}^N a_i^p \right)\right\vert^2    
Z_{\rm 1-loop}(a;m) \left\vert Z_{\rm inst}(\tau, \tau', a; m) \right\vert^2 \, , 
\fe
where the integration variables $a_i$ are constrained by $\sum_{i=1}^N a_i =0$, and $Z_{\rm 1-loop}$ gives the perturbative contributions and $Z_{\rm inst}$ leads to non-perturbative instanton contributions.  Explicitly, the perturbative term is given by
\ie \label{eq:1loop}
Z_{\rm 1-loop}(a; m) = { \prod_{i<j} a_{ij}^2  H^2(a_{ij}) \over H^{N-1}(m) \prod_{i\neq j} H(a_{ij}+m)} \,,
\fe
where $a_{ij}=a_i-a_j$, $m$ is the mass of the adjoint hypermultiplet, and $H(z)=e^{-(1+\gamma)\, z^2}G(1+iz)G(1-iz)$ with $G(x)$ being the Barnes $G$-function. The Nekrasov instanton partition function $Z_{\rm inst}$ gives non-perturbative instanton contributions, which is well understood when deformation is not turned on, i.e. $\tau'_A=0$ \cite{Nekrasov:2002qd}. For general non-vanishing $\tau'_A$, which is of interest to our consideration, results may be found in \cite{Fucito:2015ofa}, which were utilised in \cite{Paul:2022piq} in the study of the integrated correlators $\cC_{N, p}^{(i,j)}(\tau, \bar \tau)$. When $m=0$, we have the $\mathcal{N}=4$ supersymmetry, and $Z_{\rm 1-loop}(a;0)=Z_{\rm inst}(\tau, \tau', a; 0)=1$, and therefore, 
\ie \label{eq:partitionm0}
\mathcal{Z}_N(\tau, \tau_A'; 0) =& \int d^{N-1}a  \left \vert \exp\left( i \pi \tau \sum_{i=1}^N a_i^2 +i \sum_{p>2} \pi^{p/2} \tau'_p \sum_{i=1}^N a_i^p \right)\right\vert^2     \, .
\fe

In general, all the operators we consider are linear combinations of $T_{p_1, \cdots, p_n}$ that are defined in \eqref{eq:mul-tr}. 
For the localisation computation, one can insert the operator $T_{p_1, p_2, \ldots, p_n}$ by taking the derivative
\ie \label{eq:T-derivative}
 \partial_{\tau'_{p_1, p_2, \ldots, p_n}} =  \partial_{\tau'_{p_1}}\partial_{\tau'_{p_2}} \ldots \partial_{\tau'_{p_n}}\ , 
\fe
where each $\partial_{\tau'_{p_i}}$ represents inserting the operator $T_{p_i}$. It is important to note that,  as the LHS,  the RHS in the above definition should be understood as a single derivative, for example, 
\ie \label{eq:T-derive}
 \partial_{\tau'_{p_1, p_2, \ldots, p_n}} \log \mathcal{Z}_N(\tau, \tau_A'; m) = { \partial_{\tau'_{p_1, p_2, \ldots, p_n}} \mathcal{Z}_N(\tau, \tau_A'; m) \over \mathcal{Z}_N(\tau, \tau_A'; m)} =  { \partial_{\tau'_{p_1}}\partial_{\tau'_{p_2}} \ldots \partial_{\tau'_{p_n}} \mathcal{Z}_N(\tau, \tau_A'; m) \over \mathcal{Z}_N(\tau, \tau_A'; m)} \, .
 \fe

As we commented earlier, the appearance of the vectors $v_p^{i, \mu}\, , \bar{v}_p^{j, \nu}$ is due to the Gram-Schmidt procedure to resolve the mixing of operators on $S^4$ \cite{Gerchkovitz:2016gxx}.  In general, a dimension-$p$ operator on $S^4$ could mix with operators of dimensions $(p{-}2), (p{-}4)$ etc. Such mixing is resolved through the Gram–Schmidt procedure by requiring the connected two-point functions of operators with different dimensions to vanish. To simplify the Gram–Schmidt procedure, we will reorganise the operators into different towers (as well as subtowers) in the next subsection.

\subsection{Organisation of operators} \label{sec:re-org}

The localisation computation and the Gram–Schmidt procedure are greatly simplified by reorganising the operators appropriately. We will organise all superconformal primary operators as \cite{Gerchkovitz:2016gxx}
\ie \label{eq:def-opreator}
\mathcal{O}^{(i)}_{p|M} = (T_2)^p \mathcal{O}^{(i)}_{0|M}\, ,
\fe
where $(T_2)^p$ denotes the multi-trace operator $T_{2,2,\ldots, 2}$ as defined in \eqref{eq:mul-tr}, i.e. it has $p$ copies of $T_2$ (which has also been called $\mathcal{O}_2$). The operator $\mathcal{O}^{(i)}_{0|M}$, which will be defined shortly, has dimension $M$  and consequently $\mathcal{O}^{(i)}_{p|M}$ has dimension $2p+M$. To define $\mathcal{O}^{(i)}_{0|M}$, we begin by considering a general operator $T_{p_1, p_2, \ldots, p_n}$ as given in \eqref{eq:mul-tr} with all $p_i >2$ (namely, $T_2$ is excluded). These operators are ordered according to their dimensions, which are labeled as $B^{(i)}_M$ with dimension $M$ and $i$ denotes possible degeneracy. In the case of no degeneracy, we may simply omit the $i$ index. The first few examples are $B_0 = \mathbb{I}, B_3 = T_3$, $B_4 = T_4$, and $B_5 = T_5$ (there is no degeneracy for $M<6$). For $M=0$, we define $\mathcal{O}_{0|0} = B_0 = \mathbb{I}$, which is the identity operator, and then $\mathcal{O}^{(i)}_{0|M}$ with $M \geq 3$ are defined recursively as
\ie \label{eq:defOm}
\mathcal{O}^{(i)}_{0|M}  = B_M^{(i)} - \sum_{ M'\leq M} \sum_j \frac{\langle B_M^{(i)} \, , \mathcal{O}^{(j)}_{0|M'} \rangle_{S^4}}{ \langle \mathcal{O}^{(j)}_{\delta|M'}\, , \mathcal{O}^{(j)}_{0|M'} \rangle_{S^4}} \,  \mathcal{O}^{(j)}_{\delta|M'} \, , 
\fe
where we have used the definition \eqref{eq:def-opreator} for $\mathcal{O}^{(j)}_{\delta|M'}$, and the summation over $j$ is restricted to $j<i$ when $M'=M$. Here $\delta=(M-M')/2$ so that $\mathcal{O}^{(j)}_{\delta|M'}$ has the same dimension as $B^{(i)}_M$. The two-point function on $S^4$ is defined as
\ie
\langle \mathcal{O}_p\, , \mathcal{O}_q \rangle_{S^4}
= {\partial_{\tau'_p}\partial_{\tau'_q} \mathcal{Z}_N(\tau, \tau_A'; 0) \big{\vert}_{\tau_A'=0} \over \mathcal{Z}_N(\tau, 0; 0) } \, ,
\fe
with $\mathcal{Z}_N(\tau, \tau_A'; 0)$ given in \eqref{eq:partitionm0}. 
Importantly, the operators $\mathcal{O}^{(i)}_{0|M}$ constructed in this way obey the following orthogonality condition, 
\ie \label{eq:orthog1}
\langle \mathcal{O}^{(i_1)}_{0|M_1} \, , \,\,  \mathcal{O}^{(i_2)}_{0|M_2} \rangle_{S^4} = 0 \,, \quad {\rm unless}  \quad M_1 = M_2 \quad {\rm and} \quad i_1 =i_2\, ,
 \fe 
which further implies for  any $p_1, p_2$ \cite{Gerchkovitz:2016gxx}
\ie \label{eq:orthog}
\langle \mathcal{O}^{(i_1)}_{p_1|M_1} \, , \,\,  \mathcal{O}^{(i_2)}_{p_2|M_2} \rangle_{S^4} = 0 \,, \quad {\rm unless}  \quad M_1 = M_2 \quad {\rm and} \quad i_1 =i_2\, .
 \fe 
The first few examples of $\mathcal{O}^{(i)}_{0|M}$ are given by, 
\ie \label{eq:exOm}
\mathcal{O}_{0|0} =\mathbb{I}\, , \qquad \mathcal{O}_{0|3} = T_3\, , \qquad \mathcal{O}_{0|4} = T_4- {2N^2-3\over N(N^2+1)} T_{2,2}\, ,
\qquad \mathcal{O}_{0|5} = T_5- {5(N^2-2)\over N(N^2+1)} T_{2,3} \, .
\fe
For $M<6$ there is no degeneracy, and these operators actually coincide with the so-called single-particle operators $\mathcal{S}_M$ \cite{Aprile:2020uxk}, which are defined to have vanishing two-point functions with all the multi-trace operators, namely
\ie \label{eq:OmTmulti}
\langle \mathcal{S}_M \, , \,\, T_{q_1,q_2, \ldots, q_n} \rangle =0 \, , \qquad {\rm for} \qquad  n\geq 2 \, . 
\fe
There is non-trivial degeneracy when $M \geq 6$. For example, when $M=6$, we can choose $B_6^{(1)}= T_{3,3}$ and $B_6^{(2)}= T_{6}$, and we find, 
\ie \label{eq:exOm2}
\mathcal{O}_{0|6}^{(1)} &= T_{3,3} -{9\over N^2+7} T_{4,2} +\frac{3 \left(5 N^2+1\right)}{N
   \left(N^2+3\right) \left(N^2+7\right)} T_{2,2,2}\, , \cr
\mathcal{O}^{(2)}_{0|6} &= T_6 - \frac{3N^4-11N^2+80}{N(N^4+15N^2+8)}T_{3,3}- \frac{6(N^2-4)(N^2+5)}{N(N^4+15N^2+8)}T_{4,2}+ \frac{7(N^2-7)}{N^4+15N^2+8}T_{2,2,2} \, ,
\fe
where $\mathcal{O}^{(2)}_{0|6}$ is again the single-particle operator at dimension six \cite{Aprile:2020uxk}, whereas $\mathcal{O}_{0|6}^{(1)}$, which is orthogonal to $T_{2,2,2}$ and $T_{4,2}$ but not $T_{3,3}$, may be thought as the ``next-to-single-particle operator".

 In this way, we organise the operators into different towers, and within a given tower, there may be also subtowers. In particular, $\mathcal{O}^{(i)}_{p|M}$ (for any $p$) is in the $i$-th subtower of the $M$-th tower.  The operators $\mathcal{O}_{p|0}$  (and $\mathcal{O}_{p|3}$) are also known as the maximal-trace operators \cite{Paul:2022piq}, as they contain a maximal number of traces for operators with a given dimension $2p$ (and $2p+3$). The property \eqref{eq:orthog} implies that operators in different towers as well as different subtowers are orthogonal to each other on $S^4$.
Therefore, the Gram-Schmidt procedure only needs to be applied to operators within a given subtower, which greatly simplifies the procedure and consequently the localisation computation.   

\subsection{Gram-Schmidt procedure}

In this subsection, we will determine the mixing coefficients $v^{\mu}_{p|M}$ for all the operators $\mathcal{O}_{p|M}^{(i)}$ that are defined according to \eqref{eq:def-opreator}. As we have emphasised, because of the important orthogonal property \eqref{eq:orthog}, we only need to apply the Gram-Schmidt procedure for the operators in the same subtower. The mixing coefficients $v^{\mu}_{p|M}$ are determined by requiring 
\ie \label{eq:vMp}
 \left \langle  v^{\mu}_{p|M} \mathcal{O}_{\mu|M}^{(i)} \, ,  \mathcal{O}_{q|M}^{(i)}  \right \rangle_{c} =0 \, , \qquad {\rm with} \qquad q=0, 1, \ldots, p-1 \, ,
\fe
where, again, the repeated index $\mu$ is summed over, with $\mu=0,1,\ldots,p$,  and when $\mu=p$, we have $v^{p}_{p|M}=1$.  The connected two-point function is defined as
\ie \label{eq:MMpp}
 \left\langle  \mathcal{O}_{p|M}^{(i)} \, ,  \mathcal{O}_{p'|M'}^{(i')} \right\rangle_{c} & = 
\left\langle  \mathcal{O}_{p|M}^{(i)} \, ,  \mathcal{O}_{p'|M'}^{(i')} \right\rangle_{S^4} - \left\langle  \mathcal{O}_{p|M}^{(i)} \right\rangle_{S^4} \left\langle \mathcal{O}_{p'|M'}^{(i')} \right\rangle_{S^4} \, , 
\cr
&= {\partial_{{\tau'}_{p|M}^{(i)}} 
 \partial_{{\bar{\tau}}^{{}_\prime (i')}_{p'|M'}}
 \mathcal{Z}_N(\tau, \tau_A'; 0) \big{\vert}_{\tau_A'=0} \over \mathcal{Z}_N(\tau, 0; 0) }- 
{\partial_{{\tau'}_{p|M}^{(i)}} \mathcal{Z}_N(\tau, \tau_A'; 0)   \over \mathcal{Z}_N(\tau, 0; 0)  }{   \partial_{{\bar{\tau}}^{{}_\prime (i')}_{p'|M'}} \mathcal{Z}_N(\tau, \tau_A'; 0) \big{\vert}_{\tau_A'=0} \over \mathcal{Z}_N(\tau, 0; 0) } \, , 
\fe
where $\partial_{{\tau'}_{p|M}^{(i)}}$ is to insert the operator $\mathcal{O}_{p|M}^{(i)}$, which can be further expressed in terms of $T_{p_1, p_2, \ldots, p_n}$ using the definition \eqref{eq:def-opreator} and \eqref{eq:defOm} (see examples in \eqref{eq:exOm} and \eqref{eq:exOm2}).  The insertion of the operator $T_{p_1, p_2, \ldots, p_n}$ is then given by \eqref{eq:T-derivative}, as we described previously. Note by the construction of the operators $\mathcal{O}_{p|M}^{(i)}$, we have $\left\langle  \mathcal{O}_{p|M}^{(i)} \right\rangle_{S^4} = \left\langle  \mathcal{O}_{p|M}^{(i)}\, , \mathcal{O}_{0|0} \right\rangle_{S^4} =0$ if $M>0$, therefore 
\ie \label{eq:Mn0}
 \left\langle  \mathcal{O}_{p|M}^{(i)} \, ,  \mathcal{O}_{p'|M'}^{(i')} \right\rangle_{c} ={\partial_{{\tau'}_{p|M}^{(i')}} 
 \partial_{{\bar{\tau}}^{{}_\prime (i)}_{p'|M'}} \mathcal{Z}_N(\tau, \tau_A'; 0) \big{\vert}_{\tau_A'=0} \over \mathcal{Z}_N(\tau, 0; 0) } \, ,
 \fe
if $M \neq 0$ or $M' \neq 0$. When $M=M'=0$, we will need to keep both terms in \eqref{eq:MMpp}.

The set of equations \eqref{eq:vMp} uniquely determine $v^{\mu}_{p|M}$. We find the mixing coefficients $v_{p|M}^{\mu}$  for the operators $\mathcal{O}_{p|M}^{(i)}$ are in fact independent of the index $i$. Explicitly, they take the form 
\ie \label{eq:vgeneral1-app}
v_{p|M}^{\mu} = {p \choose \mu} \frac{\left( \frac{N^2+2M-1}{2}+\mu \right)_{p-\mu}}{(2\, i \, \tau_2)^{p-\mu}} \, ,
\fe
where $(x)_n$ is the Pochhammer symbol. 
Note that $\mu,p$ both start from $1$ when $M=0$, but they start from $0$ when $M>0$ \footnote{This is because for $M=0$, the first non-trivial operator is $T_2$, which according to our definition is $\mathcal{O}_{1|0}$, i.e. $M=0, p=1$}. When $M=0$ it reproduces the result given in \cite{Paul:2022piq}, with the first few examples as given below, 
\ie
v_{1|0} &= \left\{1,\quad 0, \quad 0,\quad 0,\quad \ldots \right\} \, , \cr
v_{2|0} &= \left\{- \frac{ i \left(N^2+1\right)}{2 \,  \tau_2}, \quad 1,\quad 0, \quad 0,\quad \ldots \right\}\, , \cr
v_{3|0} &= \left\{-\frac{3\left(N^2+1\right) \left(N^2+3\right)}{16\, \tau_2^2}, \quad -\frac{3\,i \left(N^2+3\right)}{4\, \tau_2},\quad 1, \quad 0,\quad \ldots \right\} \, .
\fe
Generally, for $M>0$, we have,  
\ie
v_{0|M} &= \left\{1,\quad 0, \quad 0,\quad 0,\quad \ldots \right\} \, , \cr
v_{1|M} &= \left\{- \frac{ i \left(N^2+2M-1\right)}{4 \,  \tau_2}, \quad 1,\quad 0, \quad 0,\quad \ldots \right\}\, , \cr
v_{2|M} &= \left\{-\frac{\left(N^2+2M-1\right) \left(N^2+2M+1\right)}{16\, \tau_2^2}, \quad -\frac{i \left(N^2+2M+1\right)}{2\, \tau_2},\quad 1, \quad 0,\quad \ldots \right\} \, .
\fe

We will now prove that the mixing coefficient $v^{\mu}_{p|M}$ given in \eqref{eq:vgeneral1-app} is indeed the solution to the equations \eqref{eq:vMp}. Consider the $M \neq 0$ case, for which we will use \eqref{eq:Mn0}. We first note, 
\ie \label{eq:deMMp}
\partial_{{\tau'}^{(i)}_{\mu|M}}\partial_{{\bar{\tau}}^{{}_\prime (i')}_{\nu|M'}} \mathcal{Z}_N(\tau, \tau_A'; 0) \big{\vert}_{\tau_A'=0} = \partial^{\mu}_{\tau} \, \partial^{\nu}_{\bar \tau} \,\partial_{{\tau'}^{(i)}_{0|M}}\partial_{{\bar{\tau}}^{{}_\prime (i')}_{0|M'}} \mathcal{Z}_N(\tau, \tau_A'; 0) \big{\vert}_{\tau_A'=0} \, ,
\fe
and 
\ie \label{eq:Z-beha}
\partial_{{\tau'}^{(i)}_{0|M}}\partial_{{\bar{\tau}}^{{}_\prime (i)}_{0|M}} \mathcal{Z}_N(\tau, \tau_A'; 0) \big{\vert}_{\tau_A'=0} \sim \tau_2^{-{N^2+2M-1 \over 2}} \, .
\fe
The behaviour \eqref{eq:Z-beha} can be shown as follows. 
Recall from  \eqref{eq:partitionm0}, when $m=\tau_A'=0$, the partition function is simply, 
 \ie \label{eq:MZ}
\mathcal{Z}_N(\tau, 0; 0) =& \int d^{N-1}a  \, \prod_{1\leq i<j \leq N} a_{ij}^2 \,  \exp\left( -2\pi \tau_2 \sum_{i=1}^N a_i^2  \right) \sim \tau_2^{-{N^2-1 \over 2}} \, .
 \fe
Each differential operator $\partial_{{\tau'}^{(i)}_{0|M}}$ inserts a dimension-$M$ operator $\mathcal{O}_{0|M}^{(i)}$, which gives an additional factor of $\tau_2^{- M/2}$, therefore together with \eqref{eq:MZ} we find the right behaviour  as given in \eqref{eq:Z-beha}. From \eqref{eq:deMMp} and \eqref{eq:Z-beha}, we have
\ie
& \partial_{{\tau'}^{(i)}_{\mu|M}}\partial_{{\bar{\tau}}^{{}_\prime (i)}_{q|M}} \mathcal{Z}_N(\tau, \tau_A'; 0) \big{\vert}_{\tau_A'=0} \sim  \partial^{\mu}_{\tau} \, \partial^{q}_{\bar \tau} \,  \tau_2^{-{N^2+2M-1 \over 2}}  \, ,
\fe
which can be evaluated explicitly and gives 
\ie
 \partial^{\mu}_{\tau} \, \partial^{q}_{\bar \tau} \,  \tau_2^{-{N^2+2M-1 \over 2}} 
 = \frac{(-1)^q  \, 
   \left(\frac{N^2+2M-1}{2}\right
   )_{\mu+q}}{ (-2i)^{\mu+q}} {\tau_2}^{-\frac{N^2+2 M-1}{2} -\mu-q} \, .
\fe
Using the above expression and $v_{p|M}^{\mu}$ given in \eqref{eq:vgeneral1-app}, we find, 
\ie
 v^{\mu}_{p|M} \partial_{{\tau'}^{(i)}_{\mu|M}}\partial_{{\bar{\tau}}^{{}_\prime (i)}_{q|M}} \mathcal{Z}_N(\tau, \tau_A'; 0) \big{\vert}_{\tau_A'=0}  \sim 
\frac{(-1)^q  \, \left(\frac{N^2+2M-1}{2}\right
   )_{q} \Gamma\left(p-q\right)}{ 2^{p+q} \, \Gamma
   \left(-q \right
   )} {\tau_2}^{-\frac{N^2 +2 M-1}{2} -p-q} \, , 
\fe
which vanishes identically for $0 \leq q<p$. One can similarly prove the result for the $M=0$ case.

\subsection{Normalisation of integrated correlators}

We have now appropriately organised the operators and determined the corresponding mixing coefficients, and so we will now consider the integrated correlators that are associated with the four-point functions of the following general form 
\ie \label{eq:mainC}
\langle \mathcal{O}_2\, \mathcal{O}_2\, \mathcal{O}^{(i)}_{p|M}\, \mathcal{O}^{(i')}_{p'|M'} \rangle \, .
\fe
Note that the operators $\mathcal{O}^{(i)}_{p|M}$ and $\mathcal{O}^{(i')}_{p'|M'}$ need to have the same dimension, i.e. $2p+M=2p'+M'$, therefore one may remove the dependence on $p'$. We will denote these integrated correlators as $\cC_{N,p}^{(M, M'|i,i')}(\tau, \bar \tau)$.  However, the definition of $\widehat{\cC}_{N,p}^{(M, M'|i,i')}(\tau, \bar \tau)$ following  \eqref{eq:C-hat} in general is problematic now. Following \eqref{eq:CRCh}, we would naively write, 
\ie
\cC^{(M, M'|i,i')}_{N,p}(\tau, \bar \tau) = {R^{(M, M'|i,i')}_{N,p} \over 4}\widehat{\cC}^{(M, M'|i,i')}_{N,p}(\tau, \bar \tau) \, , 
\fe
where $\widehat{\cC}^{(M, M'|i,i')}_{N,p}(\tau, \bar \tau)$ according to \eqref{eq:C-hat} would be given by
\ie \label{eq:CMMiip}
\widehat{\cC}^{(M, M'|i,i')}_{N,p}(\tau, \bar \tau) &=  \frac{v_{p|M}^{ \mu}\, \bar{v}_{p'|M'}^{ \nu}\partial_{{\tau'}^{(i)}_{\mu|M}}\partial_{{\bar{\tau}}^{{}_\prime (i')}_{\nu|M'}} \partial_m^2  \log \mathcal{Z}_N(\tau, \tau_A'; m) \big{|}_{m=\tau_A'=0}  }{ v_{p|M}^{ \mu}\, \bar{v}_{p'|M'}^{\nu}\partial_{{\tau'}^{(i)}_{\mu|M}}\partial_{{\bar{\tau}}^{{}_\prime(i')}_{\nu|M'}}  \log \mathcal{Z}_N(\tau, \tau_A'; 0) \big{|}_{\tau_A'=0} } \, ,
\fe
where again $\partial_{{\tau'}^{(i)}_{\mu|M}}$ represents inserting operator $\mathcal{O}^{(i)}_{\mu|M}$, with $\mu=0, 1, \ldots, p$ for $M>0$, and with $\mu=1, \ldots, p$ if $M=0$;  
and $R^{(M, M'|i,i')}_{N,p}$ is determined by the two-point function via
\ie \label{eq:RMMp}
\langle \mathcal{O}^{(i)}_{p|M}(x_1, Y_1)\, , \mathcal{O}^{(i')}_{p'|M'}  (x_2, Y_2)\rangle = R^{(M, M'|i,i')}_{N,p} \, \left( \frac{Y_1 \cdot Y_2}{ x_{12}^2} \right)^{2p+M} \,.
\fe
However, \eqref{eq:CMMiip} is only valid for the special cases with $M=M'$ and $i=i'$;  for other cases the denominator in the expression is zero, precisely due to  the orthogonality property \eqref{eq:orthog}. The integrated correlators $\cC_{N,p}^{(M, M'|i,i')}(\tau, \bar \tau)$ are of course  in general  non-zero and well-defined, because the prefactor $R^{(M, M'|i,i')}_{N,p} \sim \langle \mathcal{O}^{(i)}_{p|M}\, , \mathcal{O}^{(i')}_{p'|M'}  \rangle$ also vanishes when the denominator of \eqref{eq:C-hat} vanishes. Therefore, we have the issue of $0/0$ in these cases. 

To deal with this issue when $\mathcal{O}_{p|M}^{(i)}$ is not identical to $\mathcal{O}_{p'|M'}^{(i')}$, we first note the four-point function $\langle \mathcal{O}_2 \, \mathcal{O}_2 \, \mathcal{O}_{p|M}^{(i)} \, \mathcal{O}_{p'|M'}^{(i')}  \rangle$ can be written as
\ie
\langle \mathcal{O}_2 \, \mathcal{O}_2 \, \mathcal{O}_{p|M}^{(i)} \, \mathcal{O}_{p'|M'}^{(i')}  \rangle =\langle \mathcal{O}_2\, \mathcal{O}_2 \, \mathcal{O}_{p|M}^{(i)}  \,  \mathcal{O}_{p|M}^{(i)}  \rangle + \langle \mathcal{O}_2\, \mathcal{O}_2  \,  \mathcal{O}_{p|M}^{(i)} \, \mathcal{O}_d \rangle\, , 
\fe
where $\mathcal{O}_d =\mathcal{O}_{p'|M'}^{(i')}- \mathcal{O}_{p|M}^{(i)}$. Applying the above relation to the integrated correlators, we have
\ie \label{eq:C=C+C}
{\cC}^{(M, M'|i,i')}_{N,p}(\tau, \bar \tau)  = 
{R^{(M, M|i,i)}_{N,p} \over 4} \widehat{\cC}^{(M, M|i,i)}_{N,p}(\tau, \bar \tau) + {R_{\mathcal{O}_d} \over 4}\, \widehat{\cC}_{\mathcal{O}_d}(\tau, \bar \tau)\, ,
\fe
where the second term schematically denotes the integrated correlator associated with $\langle \mathcal{O}_2\, \mathcal{O}_2  \,  \mathcal{O}_{p|M}^{(i)} \, \mathcal{O}_d \rangle$. Using \eqref{eq:C-hat} from supersymmetric location, $\widehat{\cC}^{(M, M|i,i)}_{N,p}(\tau, \bar \tau)$ and $\widehat{\cC}_{\mathcal{O}_d}(\tau, \bar \tau)$ can be expressed as 
\ie
\widehat{\cC}^{(M, M|i,i)}_{N,p}(\tau, \bar \tau) &=  \frac{v_{p|M}^{ \mu}\, \bar{v}_{p|M}^{ \nu}\partial_{{\tau'}^{(i)}_{\mu|M}}\partial_{{\bar{\tau}}^{{}_\prime (i)}_{\nu|M}} \partial_m^2  \log \mathcal{Z}_N(\tau, \tau_A'; m) \big{|}_{m=\tau_A'=0}  }{ v_{p|M}^{ \mu}\, \bar{v}_{p|M}^{\nu}\partial_{{\tau'}^{(i)}_{\mu|M}}\partial_{{\bar{\tau}}^{{}_\prime(i)}_{\nu|M}}  \log \mathcal{Z}_N(\tau, \tau_A'; 0) \big{|}_{\tau_A'=0} } \, ,  \cr
\widehat{\cC}_{\mathcal{O}_d}(\tau, \bar \tau) &=  \frac{ {v}_{p|M}^{ \mu}\, \bar{v}_{\mathcal{O}_d}^{ \nu} \partial_{{\tau'}^{(i)}_{\mu|M}} \partial_{\bar{\tau}^{\prime}_{\nu|{\mathcal{O}}_d}} \partial_m^2  \log \mathcal{Z}_N(\tau, \tau_A'; m) \big{|}_{m=\tau_A'=0}  }{ {v}_{p|M}^{ \mu}\, \bar{v}_{\mathcal{O}_d}^{ \nu} \partial_{{\tau'}^{(i)}_{\mu|M}} \partial_{\bar{\tau}^{\prime}_{\nu|{\mathcal{O}}_d}}   \log \mathcal{Z}_N(\tau, \tau_A'; 0) \big{|}_{\tau_A'=0} } \, .
\fe
The mixing coefficient $v_{p|M}^{ \mu}$ is given in \eqref{eq:vgeneral1-app}, 
and we have used $\bar{v}_{\mathcal{O}_d}^{ \nu} \partial_{\bar{\tau}^{\prime}_{\nu|{\mathcal{O}}_d}}$ to schematically denote the insertion of the operator $\mathcal{O}_d$. We will not explicitly perform the Gram-Schmidt procedure on $\mathcal{O}_d$, as the details of $\bar{v}_{\mathcal{O}_d}^{ \nu}$ and $\partial_{\bar{\tau}^{\prime}_{\nu|{\mathcal{O}}_d}}$ are not needed for the following arguments.

Note neither $R^{(M, M|i,i)}_{N,p}, R_{\mathcal{O}_d}$ nor the denominators of $\widehat{\cC}^{(M, M|i,i)}_{N,p}(\tau, \bar \tau)$ and $\widehat{\cC}_{\mathcal{O}_d}(\tau, \bar \tau)$ given in the above vanish, so each term is well-defined. Importantly, because of the vanishing of the two-point function, $\langle \mathcal{O}_{p|M}^{(i)} \, ,\mathcal{O}_{p'|M'}^{(i')}  \rangle=0$, we have \ie
R^{(M, M|i,i)}_{N,p}+R_{\mathcal{O}_d} &=0 \, , 
\fe
and the denominators of $\widehat{\cC}^{(M, M|i,i)}_{N,p}(\tau, \bar \tau)$ and $\widehat{\cC}_{\mathcal{O}_d}(\tau, \bar \tau)$ are also related to each other via
\ie
v_{p|M}^{ \mu}\, \bar{v}_{p|M}^{ \nu}\partial_{{\tau'}^{(i)}_{\mu|M}}\partial_{{\bar{\tau}}^{{}_\prime(i)}_{\nu|M}}  \log \mathcal{Z}_N(\tau, \tau_A'; 0) \big{|}_{\tau_A'=0}  + {v}_{p|M}^{ \mu}\, \bar{v}_{\mathcal{O}_d}^{ \nu} \partial_{{\tau'}^{(i)}_{\mu|M}} \partial_{\bar{\tau}^{\prime}_{\nu|{\mathcal{O}}_d}}   \log \mathcal{Z}_N(\tau, \tau_A'; 0) \big{|}_{\tau_A'=0}=0 \, .
\fe
Therefore, the relation \eqref{eq:C=C+C} reduces to
\ie \label{eq:C=C+C2}
{\cC}^{(M, M'|i,i')}_{N,p}(\tau, \bar \tau)  &= {R^{(M, M|i,i)}_{N,p} \over 4} \left[ \frac{v_{p|M}^{ \mu}\, \bar{v}_{p|M}^{ \nu}\partial_{{\tau'}^{(i)}_{\mu|M}}\partial_{{\bar{\tau}}^{{}_\prime(i)}_{\nu|M}} \partial_m^2  \log \mathcal{Z}_N(\tau, \tau_A'; m) \big{|}_{m=\tau_A'=0}  }{ v_{p|M}^{ \mu}\, \bar{v}_{p|M}^{ \nu}\partial_{{\tau'}^{(i)}_{\mu|M}}\partial_{{\bar{\tau}}^{{}_\prime(i)}_{\nu|M}}  \log \mathcal{Z}_N(\tau, \tau_A'; 0) \big{|}_{\tau_A'=0} } \right. \cr
&\left. + \frac{{v}_{p|M}^{ \mu}\, \bar{v}_{\mathcal{O}_d}^{ \nu} \partial_{{\tau'}^{(i)}_{\mu|M}} \partial_{\bar{\tau}^{\prime}_{\nu|{\mathcal{O}}_d}} \partial_m^2  \log \mathcal{Z}_N(\tau, \tau_A'; m) \big{|}_{m=\tau_A'=0}  }{ v_{p|M}^{ \mu}\, \bar{v}_{p|M}^{ \nu}\partial_{{\tau'}^{(i)}_{\mu|M}}\partial_{{\bar{\tau}}^{{}_\prime(i)}_{\nu|M}}  \log \mathcal{Z}_N(\tau, \tau_A'; 0) \big{|}_{\tau_A'=0} }  \right]\, .
\fe
Using the fact that $\mathcal{O}_{p'|M'}^{(i')} =\mathcal{O}_d +  \mathcal{O}_{p|M}^{(i)}$, the numerators can be combined, which leads to
\ie \label{eq:C=C+C3}
{\cC}^{(M, M'|i,i')}_{N,p}(\tau, \bar \tau)  &= {R^{(M, M|i,i)}_{N,p} \over 4}  \frac{v_{p|M}^{ \mu}\, \bar{v}_{p'|M'}^{ \nu}\partial_{{\tau'}^{(i)}_{\mu|M}}\partial_{{\bar{\tau}}^{{}_\prime (i')}_{\nu|M'}} \partial_m^2  \log \mathcal{Z}_N(\tau, \tau_A'; m) \big{|}_{m=\tau_A'=0}  }{ v_{p|M}^{ \mu}\, \bar{v}_{p|M}^{ \nu}\partial_{{\tau'}^{(i)}_{\mu|M}}\partial_{{\bar{\tau}}^{{}_\prime(i)}_{\nu|M}}  \log \mathcal{Z}_N(\tau, \tau_A'; 0) \big{|}_{\tau_A'=0} } \, .
\fe
Importantly, there is a simple relation between $R^{(M, M|i,i)}_{N,p}$ and the denominator in \eqref{eq:C=C+C3}, 
\ie \label{eq:relationRZ}
R^{(M, M|i,i)}_{N,p} = {(4\tau_2)^{M+2p} \over (M+2p)^{2} } \, v_{p|M}^{ \mu}\, \bar{v}_{p|M}^{ \nu}\partial_{{\tau'}^{(i)}_{\mu|M}}\partial_{{\bar{\tau}}^{{}_\prime(i)}_{\nu|M}}  \log \mathcal{Z}_N(\tau, \tau_A'; 0) \big{|}_{\tau_A'=0} \, .
\fe
As $2p+M=2p'+M'$, this relation implies that in obtaining the expression for ${\cC}^{(M, M'|i,i')}_{N,p}(\tau, \bar \tau)$ one could also eliminate $\mathcal{O}_{p|M}^{(i)}$ in favour of $\mathcal{O}_{p'|M'}^{(i')}$, and lead to the same result for ${\cC}^{(M, M'|i,i')}_{N,p}(\tau, \bar \tau)$. This gives a consistency check of the argument.

One could simply use \eqref{eq:C=C+C3} to compute the integrated correlators ${\cC}^{(M, M'|i,i')}_{N,p}(\tau, \bar \tau)$. However, we find it is much more convenient and extremely useful to introduce a normalisation factor, and express the integrated correlators as \footnote{Using \eqref{eq:C=C+C3}, one could simply factor out $R^{(M, M|i,i)}_{N,p}$ and define 
\ie \label{eq:naive}
\widehat{\cC}^{(M, M'|i,i')}_{N,p}(\tau, \bar \tau)  &=   \frac{v_{p|M}^{ \mu}\, \bar{v}_{p'|M'}^{ \nu}\partial_{{\tau'}^{(i)}_{\mu|M}}\partial_{{\bar{\tau}}^{{}_\prime (i')}_{\nu|M'}} \partial_m^2  \log \mathcal{Z}_N(\tau, \tau_A'; m) \big{|}_{m=\tau_A'=0}  }{ v_{p|M}^{ \mu}\, \bar{v}_{p|M}^{ \nu}\partial_{{\tau'}^{(i)}_{\mu|M}}\partial_{{\bar{\tau}}^{{}_\prime(i)}_{\nu|M}}  \log \mathcal{Z}_N(\tau, \tau_A'; 0) \big{|}_{\tau_A'=0} }\, .
\fe
However, with this normalisation, we find that $\widehat{\cC}^{(M, M'|i,i')}_{N,p}(\tau, \bar \tau)$ does not obey a simple Laplace-difference relation. The denominator in this definition is not very natural, as $p,p'$ are not on an equal footing. Furthermore, we also find factoring out $\sqrt{R^{(M', M'|i',i')}_{N,p'} R^{(M, M|i,i)}_{N,p}}$ does not lead to a simple recursion relation either.} 
\ie \label{eq:newC}
{\cC}^{(M, M'|i,i')}_{N,p}(\tau, \bar \tau) = {\widetilde{R}^{(M, M'|i,i')}_{N,p}  \over 4 } \, {\widehat \cC_{N,p}}^{(M, M'|i,i')}(\tau, \bar \tau) \, ,
\fe
where  $\widehat{\cC}^{(M, M'|i,i')}_{N,p}(\tau, \bar \tau)$, which will be our main focus, is defined as
\ie \label{eq:newCh}
\widehat{\cC}^{(M, M'|i,i')}_{N,p}(\tau, \bar \tau) = \frac{v_{p|M}^{ \mu}\, \bar{v}_{p'|M'}^{ \nu}\partial_{{\tau'}^{(i)}_{\mu|M}}\partial_{{\bar{\tau}}^{{}_\prime (i')}_{\nu|M'}} \partial_m^2  \log \mathcal{Z}_N(\tau, \tau_A'; m) \big{|}_{m=\tau_A'=0}  }{ \mathcal{N}_M^{(i)} \,  D^{(M, M')}_{N,p} }\, ,
\fe
and the normalisation factor, which is independent of the coupling $\tau$, is given by
\ie \label{eq:Rt2}
\widetilde{R}^{(M, M'|i,i')}_{N,p} &= {R^{(M, M|i,i)}_{N,p} \, \mathcal{N}_M^{(i)} \,  D^{(M, M')}_{N,p}  \over   v_{p|M}^{ \mu}\, \bar{v}_{p|M}^{ \nu}\partial_{{\tau'}^{(i)}_{\mu|M}}\partial_{{\bar{\tau}}^{{}_\prime(i)}_{\nu|M}}  \log \mathcal{Z}_N(\tau, \tau_A'; 0) \big{|}_{\tau_A'=0} } \, ,
\fe
so that ${\cC}^{(M, M'|i,i')}_{N,p}(\tau, \bar \tau)$ agrees precisely with the expression given in \eqref{eq:C=C+C3}. As we will prove later, the integrated correlators normalised in this particular way, i.e. $\widehat{\cC}^{(M, M'|i,i')}_{N,p}(\tau, \bar \tau)$ given in \eqref{eq:newCh}, obey a universal Laplace-difference equation, which relates $\widehat{\cC}^{(M, M'|i,i')}_{N,p}(\tau, \bar \tau)$ with different charges $p$. 

Let us now explain all the ingredients as well as the motivation for the denominator that appeared in the definition of $\widehat{\cC}^{(M, M'|i,i')}_{N,p}(\tau, \bar \tau)$.  The factor $D^{(M, M')}_{N,p}$ in the denominator -- the crucial part in the definition of $\widehat{\cC}^{(M, M'|i,i')}_{N,p}$ -- is defined as follows: when $M'\neq 0$,
\ie \label{eq:DDnew}
D^{(M, M')}_{N,p} = i^{(M-M')/2} \, v_{p|M}^{\mu}\, \bar{v}_{p'|M'}^{ \nu}\partial^{\mu}_{\tau}\partial^{\nu}_{\bar{\tau}} \, \tau_2^{- {N^2+M+M'-1 \over 2} }\, , 
\fe
and when $M'= 0$,
\ie \label{eq:DDnew2}
D^{(M, 0)}_{N,p} =i^{M/2} \, {v_{p|M}^{\mu}\, \bar{v}_{p'|0}^{ \nu}\partial^{\mu}_{\tau}\partial^{\nu}_{\bar{\tau}} \, \tau_2^{- {N^2+M-1 \over 2} }  } - i^{M/2} \, {v_{p|M}^{\mu}\, \bar{v}_{p'|0}^{ \nu}\partial^{\mu}_{\tau} \, \tau_2^{- {N^2+M-1 \over 2}} \partial^{\nu}_{\bar{\tau}} \log \mathcal{Z}_N(\tau, 0; 0)   }   \, .
\fe
The expression of $D^{(M, M')}_{N,p}$ treats $p, p'$ on an equal footing (unlike the naive definition \eqref{eq:naive}), and as we will see it naturally generalises the definition \eqref{eq:CMMiip} that is valid only for $M=M', i=i'$.  The right choice of $D^{(M, M')}_{N,p}$  is crucial for $\widehat{\cC}^{(M, M'|i,i')}_{N,p}(\tau, \bar \tau)$ to satisfy a universal and simple Laplace-difference equation, which will be presented in the next section. Finally, the factor $\mathcal{N}_M^{(i)}$ is introduced so that when $M'=M$ and $i'=i$, $\widehat{\cC}^{(M, M'|i,i')}_{N,p}(\tau, \bar \tau)$ reduces to the original definition \eqref{eq:C-hat}. This factor is given by,\footnote{One could choose a different definition for $\mathcal{N}_M^{(i)}$, for example $$ 
\tau_2^{ {N^2+M+M'-1 \over 2}} \sqrt{\partial_{{\tau'}^{(i)}_{0|M}}\partial_{\bar{\tau}^{{}_\prime (i)}_{0|M}} \log \mathcal{Z}_N(\tau, \tau_A'; 0)\big{|}_{\tau_A'=0}\, \partial_{{\tau'}^{(i')}_{0|M'}}\partial_{\bar{\tau}^{{}_\prime (i')}_{0|M'}} \log \mathcal{Z}_N(\tau, \tau_A'; 0)\big{|}_{\tau_A'=0} } \,\, ,
$$ which is then symmetric in $M, M'$ and reduces to $\mathcal{N}_M^{(i)}$ when $M=M', i=i'$. However, this is not important for the purpose of obtaining the Laplace-difference equation \eqref{eq:gen-LDint} since all the $p$-dependence is contained in $D^{(M,M')}_{N,p}$. } 
\ie \label{eq:Ni}
\mathcal{N}_M^{(i)} &= \tau_2^{ {N^2+2M-1 \over 2}}  \partial_{{\tau'}^{(i)}_{0|M}}\partial_{\bar{\tau}^{{}_\prime (i)}_{0|M}} \log \mathcal{Z}_N(\tau, \tau_A'; 0)\big{|}_{\tau_A'=0}\,  \cr
&= {M^2  \over 2^{2M}} \, \tau_2^{ {N^2-1 \over 2} }  {R}^{(M, M|i,i)}_{N,0}   \, ,
\fe
 where we have used the relation \eqref{eq:relationRZ} and the fact that $v_{0|M}=1$.  We should stress that $\mathcal{N}_M^{(i)}$ is independent of $p$, therefore it plays no role for the recursion relation involving integrated correlators of different $p$'s. 

The motivation of defining $D^{(M, M')}_{N,p}$ in this particular way can be understood as follows. As in \eqref{eq:deMMp}, we first note, 
\ie \label{eq:deMMp2}
\partial_{{\tau'}^{(i)}_{\mu|M}}\partial_{{\bar{\tau}}^{{}_\prime (i')}_{\nu|M'}} \mathcal{Z}_N(\tau, \tau_A'; 0) \big{\vert}_{\tau_A'=0} =\partial^{\mu}_{\tau} \, \partial^{\nu}_{\bar \tau} \,\partial_{{\tau'}^{(i)}_{0|M}}\partial_{{\bar{\tau}}^{{}_\prime (i')}_{0|M'}} \mathcal{Z}_N(\tau, \tau_A'; 0) \big{\vert}_{\tau_A'=0} \, .
\fe
Therefore, in the definition \eqref{eq:DDnew},  we essentially replaced $\partial_{{\tau'}^{(i)}_{0|M}}\partial_{\bar{\tau}^{{}_\prime(i')}_{0|M'}}  \mathcal{Z}_N(\tau, \tau_A'; 0)\big{|}_{\tau_A'=0}$ in \eqref{eq:CMMiip} with $\tau_2^{- {N^2+M+M'-1 \over 2}}$ to avoid a vanishing denominator. Even though $\partial_{{\tau'}^{(i)}_{0|M}}\partial_{\bar{\tau}^{{}_\prime(i')}_{0|M'}}  \mathcal{Z}_N(\tau, \tau_A'; 0)\big{|}_{\tau_A'=0}$ vanishes unless $M=M', i=i'$, it is easy to see that as a function of $\tau_2$, it should behave as $\tau_2^{- {N^2+M+M'-1 \over 2}}$, following the same argument of obtaining \eqref{eq:Z-beha}. The additional factor $i^{(M-M')/2}$ is just to ensure that $D^{(M, M')}_{N,p}$ is real. The extra term in \eqref{eq:DDnew2} is because, when $M'=0$, the one-point function $\langle \mathcal{O}_{p'|0} \rangle_{S^4} \neq 0$.  We will see a similar difference between the cases of $M'=0$ and $M'\neq 0$ for the numerator of $\widehat{\cC}^{(M, M'|i,i')}_{N,p}(\tau, \bar \tau)$ when we write it out explicitly.

From the definition \eqref{eq:DDnew}, and the expression of mixing coefficients \eqref{eq:vgeneral1-app}, we can explicitly evaluate $D_{N,p}^{(M,M')}$. We find it can be expressed as 
\ie \label{eq:Dm1neq0}
D_{N,p}^{(M,M')} = \sum_{j=0}^p \sum_{k=0}^{p+\delta} \binom{p}{j} \binom{p+\delta}{k} \, (a+\delta+1+j)_{p-j} \, (a-\delta+1+k)_{p+\delta-k} \cr
\,(a+1)_j \, (a+j+1)_k \, (-1)^{j+k+\delta}\,    2^{-2p-\delta}\,  \tau_2^{-(a+1+2p+\delta)} \, , 
\fe
which simplifies to
\ie \label{eq:DD}
D_{N,p}^{(M,M')} = 2^{-2p-\delta} \,(a+1)_{p+\Theta(\delta)}  \, \Gamma \left(p+\delta-\Theta(\delta)+1 \right) \,\tau_2^{-(a+1+\delta+2p)} \, ,
\fe
where $\Theta(x)=0$ if $x \geq 0$ and $\Theta(x)=x$ if $x < 0$, and the parameters $a, \delta$ are given by 
\ie \label{eq:para}
a = \frac{N^2+M+M'-3}{2} \, , \qquad \delta = \frac{M-M'}{2} \, . 
\fe 
Using $2p+M=2p'+M'$, one can see that $D_{N,p}^{(M,M')}$ is symmetric under $M \leftrightarrow M', p \leftrightarrow p'$. Without loss of generality, one may assume $M \geq M'$ so that $\Theta(\delta)=0$.  When $M'=0$, we have, 
\ie \label{eq:DNp}
D_{N,p}^{(M,0)} =  \sum_{j=0}^p \sum_{k=1}^{p+\delta} \binom{p}{j} \binom{p+\delta}{k} \, (a+\delta+1+j)_{p-j} \, (a-\delta+1+k)_{p+\delta-k} \cr
\,(a+1)_j \, (a+j+1)_k \, (-1)^{j+k+\delta}\,  2^{-2p-\delta}\,  \tau_2^{-(a+1+2p+\delta)}
  \cr
-   \sum_{j=0}^p \sum_{k=1}^{p+\delta} \binom{p}{j} \binom{p+\delta}{k} \, (a+\delta+1+j)_{p-j} \, (a-\delta+1+k)_{p+\delta-k} \cr
\,(a+1)_j \, (a-\delta+1)_k \, (-1)^{j+k+\delta}\,  2^{-2p-\delta}\,  \tau_2^{-(a+1+2p+\delta)}\,,
\fe
where the sum over  $k$  now starts at $k=1$, as the mixing coefficient $v^{\mu}_{(p|0)}$ starts at $p=1$ and $\mu=1$. The second term in $D_{N,p}^{(M,0)}$ sums to 
\ie \label{eq:Dsecondtermsum}
 \sum_{j=0}^p \binom{p}{j}  \, (a+\delta+1+j)_{p-j} \, (a-\delta+1)_{p+\delta} 
\,(a+1)_j \, (-1)^{j+\delta}\,  2^{-2p-\delta}\,  \tau_2^{-(a+1+2p+\delta)} \, ,
\fe
which is in fact equivalent to the $k=0$ case of the first term. Therefore we can combine these two terms in \eqref{eq:DNp}, which leads to the same expression as \eqref{eq:Dm1neq0}. We see that with the extra term in the definition of $D_{N,p}^{(M,0)}$, the final expression for $D_{N,p}^{(M,0)}$ is simply  $D_{N,p}^{(M,M')}$ given in \eqref{eq:DD} with $M' \rightarrow 0$.

Using the relation \eqref{eq:relationRZ} and the expressions for $\mathcal{N}_N^{(i)}$  and $D_{N,p}^{(M,M')}$ as given in \eqref{eq:Ni} and \eqref{eq:DD}, we can simplify the normalisation factor $\widetilde{R}^{(M, M'|i,i')}_{N,p}$ as given in \eqref{eq:Rt2}, and find
\ie \label{eq:Rt1}
\widetilde{R}^{(M, M'|i,i')}_{N,p}
 = {2^{2p+ \delta} \, M^2  \over (M+2p)^{2} }    \,   (a+1)_{p+\Theta(\delta)}  \, \Gamma \left(p+\delta-\Theta(\delta)+1 \right)\,  {R}^{(M, M|i,i)}_{N,0}  \, .
\fe
Note by construction, when $M=M'$ and $i=i'$, $\widetilde{R}^{(M, M'|i,i')}_{N,p}$ reduces to ${R}^{(M, M|i,i)}_{N,p}$.

From now on, we will mainly focus on  ${\widehat \cC_{N,p}}^{(M, M'|i,i')}(\tau, \bar \tau)$ as defined in \eqref{eq:newCh}. In the next section, we will prove that, with this crucial normalisation factor $\widetilde{R}^{(M, M'|i,i')}_{N,p}$, ${\widehat \cC_{N,p}}^{(M, M'|i,i')}(\tau, \bar \tau)$ obeys a universal Laplace-difference equation that gives a three-term recursion relation relating the integrated correlators of different charges $p$.

\section{Laplace-difference equation} \label{sec:single}

In this section, we will show that the normalised integrated correlators $\widehat{\cC}^{(M, M'|i,i')}_{N; p}(\tau, \bar \tau)$ as defined in \eqref{eq:newCh} satisfy the following universal Laplace-difference equation, 
\ie \label{eq:gen-LDu}
\Delta_{\tau}\, {\widehat \cC_{N,p}}^{(M, M'|i, i')} & (\tau, \bar \tau)  =\, \left(p+1+ \delta  \right)\left(p+ a+1 \right) {\widehat \cC_{N,p+1}}^{(M, M'|i,i')}(\tau, \bar \tau) + p\left(p+ a +\delta \right) {\widehat \cC_{N,p-1}}^{(M, M'|i,i')}(\tau, \bar \tau) \cr
& - \left[ 2p\left( p+ a\right) + (2p+a+1)(\delta+1) \right] {\widehat \cC_{N,p}}^{(M, M'|i,i')}(\tau, \bar \tau) -4 \, \delta_{M,M'} \delta_{i, i'} \cC_{N,1}^{(0,0)}(\tau, \bar \tau) \, ,
\fe
where the parameters $a, \delta$ are given in \eqref{eq:para}, 
and $\Delta_\tau = 4 \tau_2^2 \partial_{\tau}\partial_{\bar \tau}$.  The ``source term" $\cC_{N,1}^{(0,0)}(\tau, \bar \tau)$, which only appears when $M=M'$ and $i=i'$, is given in \eqref{eq:2222} (now it is written in our unified notation). It also obeys a Laplace-difference equation \eqref{eq:N-LG} and  is known exactly for any $N$ and $\tau$ \cite{Dorigoni:2021bvj, Dorigoni:2021guq}. It is important to note that because ${\widehat \cC_{N,p}}^{(M, M'|i, i')}  (\tau, \bar \tau)=0$ when $p<0$, \eqref{eq:gen-LDu} is a recursion relation (rather than a differential equation) that determines ${\widehat \cC_{N,p}}^{(M, M'|i, i')} (\tau, \bar \tau)$ for any $p$ in terms of the initial data ${\widehat \cC_{N,0}}^{(M, M'|i, i')}  (\tau, \bar \tau)$. In the special case $M=M'=0$ (the operators were called maximal-trace operators with even dimensions), our Laplace-difference equation \eqref{eq:gen-LDu} reproduces what has been found in \cite{Paul:2022piq}. We will therefore mainly focus on $M>0$ in the cases with $M=M', i=i'$.  It is intriguing to note that the $N$-dependence of the recursion only appears in the parameter $a$ (given in \eqref{eq:para}), which has the symmetry of $N^2 \leftrightarrow M+M'$.
 
Interestingly, the difference of integrated correlators,
\ie \label{eq:ddHu}
\widehat{ \mathcal{H}}_{N,p}^{(M, M'|i, i')} & (\tau, \bar \tau) =  {\widehat \cC_{N,p}}^{(M, M'|i, i')}(\tau, \bar \tau) -  {\widehat \cC_{N,p-1}}^{(M, M'|i, i')}(\tau, \bar \tau)\, ,
\fe also satisfies a three-term recursion relation, 
\ie \label{eq:LDHu}
 \Delta_{\tau} \widehat{ \mathcal{H}}_{N,p}^{(M, M'|i, i')}(\tau, \bar \tau)=& \, (p+1+\delta) \left( p+a+1 \right) \widehat{ \mathcal{H}}_{N,p+1}^{(M, M'|i, i')}(\tau, \bar \tau)+ \, (p-1) \left(p+a+\delta-1 \right)\widehat{ \mathcal{H}}_{N,p-1}^{(M, M'|i, i')}(\tau, \bar \tau) \cr 
&- \left[(p+a)(p+\delta)+p(p+a+\delta)\right] \widehat{ \mathcal{H}}_{N,p}^{(M, M'|i, i')}(\tau, \bar \tau)\, .
\fe
Note the above recursion relation for $\widehat{ \mathcal{H}}_{N,p}^{(M, M'|i, i')}(\tau, \bar \tau)$  is valid only for $p\geq 1$.
It is worth emphasising that it is non-trivial that a three-term recurrence relation exists for the difference $\widehat{\mathcal{H}}_{N,p}^{(M, M'|i, i')}(\tau, \bar \tau)$.

In the following subsections, we will prove the Laplace-difference equation \eqref{eq:gen-LDu}. 

\subsection{Cases with $M=M', i=i'$}

In this subsection, we consider the cases with $M=M'$ and $i=i'$.  As we mentioned we will focus on $M>0$ with general $M$ here. For these particular cases, by construction the denominator in \eqref{eq:newCh} reduces to 
\ie
\mathcal{N}_M^{(i)} \,  D^{(M, M)}_{N,p} = v_{p|M}^{ \mu}\, \bar{v}_{p|M}^{ \nu}\partial_{{\tau'}^{(i)}_{\mu|M}}\partial_{{\bar{\tau}}^{{}_\prime(i)}_{\nu|M}}  \log \mathcal{Z}_N(\tau, \tau_A'; m) \big{|}_{m=\tau_A'=0} \, ,
\fe
so we have
\ie 
\widehat{\cC}^{(M, M|i,i)}_{N,p}(\tau, \bar \tau) = \frac{v_{p|M}^{ \mu}\, \bar{v}_{p|M}^{ \nu}\partial_{{\tau'}^{(i)}_{\mu|M}}\partial_{{\bar{\tau}}^{{}_\prime(i)}_{\nu|M}} \partial_m^2  \log \mathcal{Z}_N(\tau, \tau_A'; m) \big{|}_{m=\tau_A'=0}  }{ v_{p|M}^{ \mu}\, \bar{v}_{p|M}^{ \nu}\partial_{{\tau'}^{(i)}_{\mu|M}}\partial_{{\bar{\tau}}^{{}_\prime(i)}_{\nu|M}}  \log \mathcal{Z}_N(\tau, \tau_A'; 0) \big{|}_{\tau_A'=0} }\, . 
\fe
As $\mathcal{Z}_N(\tau, \tau_A'; m)$ is an even function of $m$, we have
\ie
\partial_m^2 \log \mathcal{Z}_N(\tau, \tau_A'; m) \big{|}_{m=0} ={\partial_m^2  \mathcal{Z}_N(\tau, \tau_A'; m) \big{|}_{m=0}  \over \mathcal{Z}_N(\tau, \tau_A'; 0)} \, ,
\fe
and from \eqref{eq:orthog}, we have
\ie \label{eq:van1}
\partial_{{\tau'}^{(i)}_{\mu|M}} \mathcal{Z}_N(\tau, \tau_A'; 0)\big{|}_{\tau_A'=0} =\partial_{{\bar{\tau}}^{{}_\prime(i)}_{\nu|M}}   \mathcal{Z}_N(\tau, \tau_A'; 0) \big{|}_{\tau_A'=0}   =0\, .
\fe
We therefore find
\ie  \label{eq:i=i}
\widehat{\cC}^{(M, M|i,i)}_{N,p}(\tau, \bar \tau) = \frac{v_{p|M}^{ \mu}\, \bar{v}_{p|M}^{ \nu}\partial_{{\tau'}^{(i)}_{\mu|M}}\partial_{{\bar{\tau}}^{{}_\prime(i)}_{\nu|M}} \partial_m^2  \mathcal{Z}_N(\tau, \tau_A'; m) \big{|}_{m=\tau_A'=0}  }{ v_{p|M}^{ \mu}\, \bar{v}_{p|M}^{ \nu}\partial_{{\tau'}^{(i)}_{\mu|M}}\partial_{{\bar{\tau}}^{{}_\prime(i)}_{\nu|M}}  \mathcal{Z}_N(\tau, \tau_A'; 0) \big{|}_{\tau_A'=0} } - {\partial_m^2  \mathcal{Z}_N(\tau, 0; m) \big{|}_{m=0} \over \mathcal{Z}_N(\tau, 0; 0)}\, . 
\fe

Let us first consider the second term in the above equation. The laplacian acting on it leads to
\ie
\Delta_{\tau} {\partial_m^2  \mathcal{Z}_N(\tau, 0; m) \big{|}_{m=0} \over \mathcal{Z}_N(\tau, 0; 0) } = 4 \, \cC_{N,1}^{(0,0)}(\tau, \bar \tau)\, .
\fe
Using this fact, one can see that this term alone obeys the Laplace-difference equation \eqref{eq:gen-LDu}.  Therefore, from now on we will only focus on the first term of \eqref{eq:i=i}, which we will denote as $\widehat{\cC'}^{(M, M|i,i)}_{N,p}(\tau, \bar \tau)$ and quote as below, 
\ie \label{eq:Cpp}
\widehat{\cC'}^{(M, M|i,i)}_{N,p}(\tau, \bar \tau)  = \frac{v_{p|M}^{ \mu}\, \bar{v}_{p|M}^{ \nu}\partial_{{\tau'}^{(i)}_{\mu|M}}\partial_{{\bar{\tau}}^{{}_\prime(i)}_{\nu|M}} \partial_m^2  \mathcal{Z}_N(\tau, \tau_A'; m) \big{|}_{m=\tau_A'=0}  }{ v_{p|M}^{ \mu}\, \bar{v}_{p|M}^{ \nu}\partial_{{\tau'}^{(i)}_{\mu|M}}\partial_{{\bar{\tau}}^{{}_\prime(i)}_{\nu|M}}  \mathcal{Z}_N(\tau, \tau_A'; 0) \big{|}_{\tau_A'=0} } \, .
\fe
Since the source term $-4 \, \cC_{N,1}^{(0,0)}(\tau, \bar \tau)$ arises from the second term of \eqref{eq:i=i} as we just showed, we will prove that $\widehat{\cC'}^{(M, M|i,i)}_{N,p}(\tau, \bar \tau)$ obeys  \eqref{eq:gen-LDu} without the source term,  namely
\ie \label{eq:gen-LDup}
\Delta_{\tau}\, \widehat{\cC'}^{(M, M|i,i)}_{N,p} & (\tau, \bar \tau)  =\, \left(p+1+ \delta  \right)\left(p+ a+1 \right) \widehat{\cC'}^{(M, M|i,i)}_{N,p+1}(\tau, \bar \tau)  + p\left(p+ a +\delta \right) \widehat{\cC'}^{(M, M|i,i)}_{N,p-1}(\tau, \bar \tau) \cr
& -\, \left[ 2p\left( p+ a\right) + (2p+a+1)(\delta+1) \right] \widehat{\cC'}^{(M, M|i,i)}_{N,p}(\tau, \bar \tau)   \, ,
\fe
with $\delta=0$ due to $M=M'$. In the next subsection, we will in fact see that, for the integrated correlators with $M\neq M'$ or $M= M'$ but $i\neq i'$, what we will need to prove is essentially the same as \eqref{eq:gen-LDup}  (with $\delta \neq 0$ if $M\neq M'$).

\subsection{Cases with $M \neq M'$ or $i \neq i'$}

We now consider the cases with $M\neq M'$, i.e. the operators are in different towers, or the cases with $M=M'$ but $i \neq i'$, i.e. the operators are in different subtowers. For all these cases, because the two-point functions vanish, we need to use the refined definition \eqref{eq:newCh} for $\widehat{\cC}^{(M, M'|i,i')}_{N,p}(\tau, \bar \tau)$.  We will first consider both $M$ and $M'$ are not $0$. Besides \eqref{eq:van1}, we also have the relation
\ie
\partial_{{\tau'}^{(i)}_{\mu|M}}\partial_{{\bar{\tau}}^{{}_\prime (i')}_{\nu|M'}}   \mathcal{Z}_N(\tau, \tau_A'; ) \big{|}_{\tau_A'=0}  =0\, ,
\fe
when the operators are in different (sub)towers. Therefore, unlike \eqref{eq:i=i}, $\widehat{\cC}^{(M, M|i,i')}_{N,p}(\tau, \bar \tau)$ has no analogical second term, and so
\ie \label{eq:MnMp}
\widehat{\cC}^{(M, M'|i,i')}_{N,p}(\tau, \bar \tau) = \frac{v_{p|M}^{ \mu}\, \bar{v}_{p'|M'}^{ \nu}\partial_{{\tau'}^{(i)}_{\mu|M}}\partial_{{\bar{\tau}}^{{}_\prime (i')}_{\nu|M'}} \partial_m^2  \mathcal{Z}_N(\tau, \tau_A'; m) \big{|}_{m=\tau_A'=0}  }{ \mathcal{N}_{M}^{(i)} \,  D^{(M, M')}_{N,p}  }\, .
\fe
 Therefore in these cases, there is no source term $-4 \, \cC_{N,1}^{(0,0)}(\tau, \bar \tau)$, consistent with the Laplace-difference equation \eqref{eq:gen-LDu}.

When $M'=0$, the relation \eqref{eq:van1} is not valid anymore. In particular, we have 
\ie \label{eq:dZ0}
\partial_{{\tau'}^{(1)}_{\mu|0}} \mathcal{Z}_N(\tau, \tau_A'; 0)\big{|}_{\tau_A'=0} =\partial^{\mu}_{\tau} \mathcal{Z}_N(\tau, 0; 0) \neq 0\, .
\fe
Here and in the following, because there is no degeneracy when $M'=0$, we set the degeneracy index $i$ as $1$. Equation \eqref{eq:dZ0}  implies that there is an additional term in $\widehat{\cC}^{(M, 0|i,1)}_{N,p}(\tau, \bar \tau) $, 
\ie \label{eq:Mp=0}
\widehat{\cC}^{(M, 0|i,1)}_{N,p}(\tau, \bar \tau) &= \frac{v_{p|M}^{ \mu}\, \bar{v}_{p'|0}^{ \nu}\partial_{{\tau'}^{(i)}_{\mu|M}}\partial_{\bar{\tau}^{{}_\prime (1)}_{\nu|0}} \partial_m^2  \mathcal{Z}_N(\tau, \tau_A'; m) \big{|}_{m=\tau_A'=0}  }{ \mathcal{N}_M^{(i)} \,  D^{(M, 0)}_{N,p} }\cr
&- \, \frac{v_{p|M}^{ \mu}\, \bar{v}_{p'|0}^{ \nu}  \partial_{{\tau'}^{(i)}_{\mu|M}} \partial_m^2  \mathcal{Z}_N(\tau, \tau_A'; m) \big{|}_{m=\tau_A'=0} \, \partial_{\bar{\tau}^{{}_\prime (1)}_{\nu|0}} \mathcal{Z}_N(\tau, 0; 0)  }{ \mathcal{N}_M^{(i)} \,  D^{(M, 0)}_{N,p} \,\mathcal{Z}_N(\tau, 0; 0) } \, .
\fe
This important difference between  \eqref{eq:MnMp} and \eqref{eq:Mp=0} is similar to the definition \eqref{eq:DDnew2} for the normalisation factor in the case of $M'=0$, as we emphasised earlier. 

In the next subsection, using these explicit expressions of $\widehat{\cC}^{(M, M'|i,i')}_{N,p}(\tau, \bar \tau)$ we have obtained in the above analysis, we will prove that $\widehat{\cC}^{(M, M'|i,i')}_{N,p}(\tau, \bar \tau)$ satisfies the universal Laplace-difference equation \eqref{eq:gen-LDu}.

\subsection{Proof of Laplace-difference equation}\label{sec:proof}

We will now prove the Laplace-difference equation \eqref{eq:gen-LDu}. Just as in \eqref{eq:deMMp}, we have,  
\ie
\partial_{{\tau'}^{(i)}_{\mu|M}}\partial_{\bar{\tau}^{{}_\prime (i')}_{\nu|M'}} \partial_m^2  \mathcal{Z}_N(\tau, \tau_A'; m) \big{|}_{m=\tau_A'=0} = \partial_{\tau}^{\mu}\partial_{\bar \tau}^{\nu} \partial_{{\tau'}^{(i)}_{0|M}}\partial_{\bar{\tau}^{{}_\prime (i')}_{0|M'}} \partial_m^2  \mathcal{Z}_N(\tau, \tau_A'; m) \big{|}_{m=\tau_A'=0} \, ,
\fe
therefore  $\widehat{\cC}^{(M, M'|i,i')}_{N,p}(\tau, \bar \tau)$ can be written in terms of $\partial_{{\tau'}^{(i)}_{0|M}}\partial_{\bar{\tau}^{{}_\prime (i')}_{0|M'}} \partial_m^2  \mathcal{Z}_N(\tau, \tau_A'; m) \big{|}_{m=\tau_A'=0}$ with derivatives with respect to $\tau$ and $\bar{\tau}$ acting on it. To prove the Laplace-difference equation \eqref{eq:gen-LDu}, we will collect all the terms with the same number of derivatives of $\tau$ and $\bar{\tau}$, and show that the corresponding coefficients vanish. Therefore we find that the  Laplace-difference equation does not rely on the precise form of $\partial_{{\tau'}^{(i)}_{0|M}}\partial_{\bar{\tau}^{{}_\prime (i')}_{0|M'}} \partial_m^2  \mathcal{Z}_N(\tau, \tau_A'; m) \big{|}_{m=\tau_A'=0}$. Furthermore, as we discussed in the previous subsections, because there is some difference between the cases of $M'=0$ and the cases of $M'>0$, we will consider them separately. Without losing generality, we will assume $M \geq M'$ in the following discussion.

\subsubsection{Cases with $M' > 0$}

We begin with the case $M' > 0$.  Using the result of $D_{N,p}^{(M,M')}$ as given in \eqref{eq:DD} and the mixing coefficients \eqref{eq:vgeneral1-app}, we can express $\widehat{\cC}^{(M, M'|i,i')}_{N,p}(\tau, \bar \tau)$ as 
\ie \label{Mpn=0}
 & \widehat{\cC}^{(M, M'|i,i')}_{N,p}  (\tau, \bar \tau) = \frac{1}{ \mathcal{N}_M^{(i)} \, (a+1)_p (p+\delta)!} \sum_{j=0}^p \sum_{k=0}^{p+\delta} \,\binom{p}{j} \binom{p+\delta}{k} \, (a+\delta+j+1)_{p-j} \cr
& (a-\delta+k+1)_{p+\delta-k} \,
(-1)^{k-\delta}\, 2^{j+k}\, i^{j+k-\delta} \,\tau_2^{j+k+a+1}\, \partial_\tau^j \, \partial_{\bar{\tau}}^k \, \partial_{{\tau'}^{(i)}_{0|M}}\, \partial_{\bar{\tau}^{{}_\prime (i')}_{0|M'}} \, \partial_m^2 \mathcal{Z}_N(\tau, \tau_A'; m)\big{|}_{m=\tau_A'=0} \, .
\fe
As we commented above, the validity of the Laplace-difference equation is actually independent of the precise form of $\partial_{{\tau'}^{(i)}_{0|M}}\, \partial_{\bar{\tau}^{{}_\prime (i')}_{0|M'}} \, \partial_m^2 \mathcal{Z}_N(\tau, \tau_A'; m) \big{|}_{m=\tau_A'=0}$. It holds separately for every 
\ie
\partial_\tau^j \, \partial_{\bar{\tau}}^k \, \partial_{{\tau'}^{(i)}_{0|M}}\, \partial_{\bar{\tau}^{{}_\prime (i')}_{0|M'}} \, \partial_m^2 \mathcal{Z}_N(\tau, \tau_A'; m) \big{|}_{m=\tau_A'=0}  \, .
\fe
We therefore isolate the $j$ and $k$-th derivative 
\ie
\widehat{\cC}^{(M, M'|i,i')}_{N,p} & (\tau, \bar \tau) \big{\vert}_{j,k} =  \frac{(-1)^{k-\delta}\, 2^{j+k} \,i^{j+k-\delta}\, \tau_2^{j+k+a+1}}{\mathcal{N}_M^{(i)}\, (a+1)_p \,(p+\delta)!} \binom{p}{j} \binom{p+\delta}{k}  \cr 
& (a+\delta+j+1)_{p-j} \,(a-\delta+k+1)_{p+\delta-k} \,
\partial_\tau^j \, \partial_{\bar{\tau}}^k \, \partial_{{\tau'}^{(i)}_{0|M}}\, \partial_{\bar{\tau}^{{}_\prime (i')}_{0|M'}} \, \partial_m^2 \mathcal{Z}_N(\tau, \tau_A'; m)\big{|}_{m=\tau_A'=0} \, .
\fe
The above expression implies
\ie
\widehat{\cC}^{(M, M'|i,i')}_{N,p+1}(\tau, \bar \tau)\big{\vert}_{j,k} = \frac{(p+1)(a+1+p+\delta)}{(p+1-j)(p+\delta+1-k)} \widehat{\cC}^{(M, M'|i,i')}_{(N,p)}(\tau, \bar \tau)\big{\vert}_{j,k} \, ,
\fe
and 
\ie
\widehat{\cC}^{(M, M'|i,i')}_{N,p-1}(\tau, \bar \tau) \big{\vert}_{j,k} = \frac{(p-j)(p+\delta-k)}{p(a+p+\delta)} \widehat{\cC}^{(M, M'|i,i')}_{(N,p)}(\tau, \bar \tau) \big{\vert}_{j,k} \, .
\fe
Taking the laplacian $\Delta_\tau = 4 \tau_2^2 \partial_{\tau}\partial_{\bar \tau}$ and collecting the $j$ and $k$-th derivatives, we find 
\ie
\Delta_\tau \widehat{\cC}^{(M, M'|i,i')}_{N,p}(\tau, \bar \tau) \big{\vert}_{j,k}  = \left((a+j+k) (a+j+k+1)+ \frac{j \,(a+j+k) (a+\delta+j)}{p-j+1} \right. \cr 
   \left. +\frac{k \,(a+j+k)
   (a-\delta+k)}{p+\delta+1-k}  +\frac{j \,k \,(a+\delta+j)
   (a-\delta+k)}{(p-j+1) (p+\delta+1-k)}\right) \widehat{\cC}^{(M, M'|i,i')}_{N,p}(\tau, \bar \tau) \big{\vert}_{j,k}  \, .
\fe

Let us now consider the following combination that is relevant for the Laplace-difference equation \eqref{eq:gen-LDu}, 
\ie \label{eq:Lap0}
\Delta_{\tau}\,& {\widehat \cC_{N,p}}^{(M, M'|i, i')}  (\tau, \bar \tau)\big{\vert}_{j,k} + \left[ 2p\left( p+ a\right) + (2p+a+1)(\delta+1) \right] {\widehat \cC_{N,p}}^{(M, M'|i,i')}(\tau, \bar \tau)\big{\vert}_{j,k} \cr
 &-\, \left(p+1+ \delta  \right)\left(p+ a+1 \right) {\widehat \cC_{N,p+1}}^{(M, M'|i,i')}(\tau, \bar \tau) \big{\vert}_{j,k} - p\left(p+ a +\delta \right) {\widehat \cC_{N,p-1}}^{(M, M'|i,i')}(\tau, \bar \tau)\big{\vert}_{j,k}  \, .
\fe
Using the above results, \eqref{eq:Lap0} becomes 
\ie
& \Big[ (a+j+k) (a+j+k+1)+ \frac{j (a+j+k) (a+\delta+j)}{p-j+1} +\frac{k (a+j+k)
   (a-\delta+k)}{p+\delta+1-k} \cr 
 & + \frac{j k (a+\delta+j)
   (a-\delta+k)}{(p-j+1) (p+\delta+1-k)}    - \frac{(p+1)(p+\delta+1)(a+p+\delta+1)(a+p+1)}{(p+1-j)(p+\delta+1-k)} \cr
& + (p-j)(p+\delta-k)      -(2p(p+a)+(2p+a+1)(\delta+1)) \Big] \,  {\widehat \cC}_{N,p}^{(M, M'|i, i')} (\tau, \bar \tau)\big{\vert}_{j,k} \,.
\fe
It is straightforward to check that the factor in front of ${\widehat \cC}_{N,p}^{(M, M'|i, i')} (\tau, \bar \tau)\big{\vert}_{j,k}$ in the above expression identically evaluates to zero. 

Clearly the above discussion also applies to the special case $M=M'$ and $i=i'$, namely $\widehat{\cC'}^{(M, M|i,i)}_{N,p}(\tau, \bar \tau)$ defined in \eqref{eq:Cpp} and the relation \eqref{eq:gen-LDup} it satisfies. This proves the Laplace-difference equation for the case $M'>0$. 

\subsubsection{Cases with $M'=0$}

The proof of the Laplace-difference equation for the case $M'=0$ is similar.  When $M'=0$, analogous to $D_{N,p}^{(M,0)}$ in the denominator in \eqref{eq:DDnew2},  the numerator of $\widehat{\cC}^{(M, 0|i,1)}_{N,p}(\tau, \bar \tau)$ also has an extra term, as given in \eqref{eq:Mp=0}. Explicitly, $\widehat{\cC}^{(M, 0|i,1)}_{N,p}(\tau, \bar \tau)$ can be expressed as 
\ie
& \widehat{\cC}^{(M, 0|i,1)}_{N,p} (\tau, \bar \tau) = \frac{1}{\mathcal{N}_M^{(i)} \,(a+1)_p (p+\delta)!} \sum_{j=0}^p \sum_{k=1}^{p+\delta} \binom{p}{j} \binom{p+\delta}{k} \, (a+\delta+1+j)_{p-j} \cr
& (a-\delta+1+k)_{p+\delta-k} (-1)^{k-\delta} \,  2^{j+k} \, i^{j+k-\delta} \, \tau_2^{j+k+a+1} \, \partial_\tau^j \, \partial_{\bar{\tau}}^k \, \partial_{{\tau'}^{(i)}_{0|M}}\, \partial_{\bar{\tau}^{{}_\prime (1)}_{0|M'}} \, \partial_m^2 \mathcal{Z}_N(\tau, \tau_A'; m) \big{|}_{m=\tau_A'=0} \cr
&-\frac{1}{\mathcal{N}_M^{(i)} (a+1)_p (p+\delta)!} \sum_{j=0}^p \sum_{k=1}^{p+\delta} \binom{p}{j} \binom{p+\delta}{k} \, (a+\delta+1+j)_{p-j} \cr
&(a-\delta+1+k)_{p+\delta-k}  (a-\delta+1)_k (-1)^{k-\delta} \, 2^{j} \, i^{j-\delta} \, \tau_2^{j+a+1} \, \partial_\tau^j \, \partial_{{\tau'}^{(i)}_{0|M}}\, \partial_{\bar{\tau}^{{}_\prime (1)}_{0|M'}} \, \partial_m^2 \mathcal{Z}_N(\tau, \tau_A'; m)\big{|}_{m=\tau_A'=0} \, .
\fe 
Similarly to  $D_{N,p}^{(M,0)}$ in \eqref{eq:Dsecondtermsum}, the second term in $\widehat{\cC}^{(M, 0|i,1)}_{N,p} (\tau, \bar \tau)$ can be simplified and sums to 
\ie
 \frac{1}{\mathcal{N}_M^{(i)}  \, (a+1)_p (p+\delta)!}&  \sum_{j=0}^p \binom{p}{j}  (a+\delta+1+j)_{p-j} (a-\delta+1)_{p+\delta}
\cr 
&  (-1)^{-\delta} \,2^{j} \,i^{j-\delta}\, \tau_2^{j+a+1} \, \partial_\tau^j \, \partial_{{\tau'}^{(i)}_{0|M}}\, \partial_{\bar{\tau}^{{}_\prime (1)}_{0|M'}} \, \partial_m^2 \mathcal{Z}_N(\tau, \tau_A'; m) \big{|}_{m=\tau_A'=0} \, , 
\fe
which is equivalent to the $k=0$ case of the first term, and so the integrated correlator can be expressed as 
\ie
& \widehat{\cC}^{(M, 0|i,1)}_{N,p} (\tau, \bar \tau) = \frac{1}{\mathcal{N}_M^{(i)}\, (a+1)_p (p+\delta)!} \sum_{j=0}^p \sum_{k=0}^{p+\delta} \binom{p}{j} \binom{p+\delta}{k} \, (a+\delta+1+j)_{p-j} \cr
&~ (a-\delta+1+k)_{p+\delta-k} \, (-1)^{k-\delta} \,2^{j+k}\, i^{j+k-\delta} \,\tau_2^{j+k+a+1} \,\partial_\tau^j \, \partial_{\bar{\tau}}^k \, \partial_{{\tau'}^{(i)}_{0|M}}\, \partial_{\bar{\tau}^{{}_\prime (1)}_{0|M'}} \, \partial_m^2 \mathcal{Z}_N(\tau, \tau_A'; m) \big{|}_{m=\tau_A'=0} \, .
\fe
We see that $\widehat{\cC}^{(M, 0|i,1)}_{N,p}(\tau, \bar \tau)$ takes the same form as that for $M' \neq 0$ as given in \eqref{Mpn=0}. Therefore, the proof of the Laplace-difference equation for the $M' \neq 0$ case applies to $M'=0$ as well. This completes the proof of the Laplace-difference equation \eqref{eq:gen-LDu}.

\section{Examples of integrated correlators} \label{sec:local}

In this section, we will give some explicit results for certain examples of the integrated correlators $\widehat{\cC}^{(M, M'|i,i')}_{N,p}(\tau, \bar \tau)$ and the corresponding prefactors $\widetilde{R}_{N;p}^{(M, M'|i,i')}$. 
Here we will mainly focus on the perturbative contributions, i.e. the zero-instanton sector. However, as we will comment on further, with some appropriate assumptions of $SL(2, \mathbb{Z})$ modular properties of the integrated correlators, the perturbative contributions actually uniquely determine the full integrated correlators. For the perturbative contributions, we can simply omit the instantonic contribution $Z_{\rm inst}$ in the partition function \eqref{eq:partition}, and the small-$m$ expansion of $Z_{\rm 1-loop}$ can be straightforwardly evaluated  by using \cite{Russo:2013kea},
\ie
\partial_m^2 \log {   H^2(a) \over  H(a{+}m)H(a{-}m)} \Big{\vert}_{m=0} &= - 4 \int_0^{\infty} dw  {w \left( \cos (2 w a) - 1 \right) \over \sinh^2(w) } \cr
&= -4 \sum_{\ell=1}^{\infty} (-1)^\ell \, (2 \ell + 1)\, \zeta(2 \ell + 1) \, a^{\ell + 1}\, .
\fe
Furthermore, the derivative $\partial_{\tau'_p}$ is to insert $i \pi^{p/2} \sum_i a_i^p$. Therefore, in the perturbation, the integrals that we need to compute are gaussian integrals of the form 
\ie
 \int d^{N}a \, \delta\left( \sum_{i=1}^N a_i \right)   \exp\left( - 2\pi  \tau_2 \sum_{i=1}^N a_i^2  \right)   \left( \prod_{i<j} a^2_{ij}  \right) \left( \prod_{j}  \sum_{i=1}^N a_i^{p_j}  \right) \, .
\fe
It is straightforward to compute these integrals explicitly, at least up to some finite orders. To illustrate the structures, below we will provide the first few orders in perturbation for the integrated correlators $\widehat{\cC}^{(M,M'|i,i')}_{N,p}$ with $M, M' \leq 5$, and as we mentioned they coincide with the single-particle operators when $p=0$. Since there is no degeneracy for these cases, we will simply drop the indices $i$ and $i'$ for all these examples.

\subsection{Integrated correlators with $M=M'$}

We begin with the integrated correlators with $M=M'$. The special case $M=M'=0$ has been considered in \cite{Paul:2022piq}, so we will only consider those with $M>0$, and the results are listed as follows:

\subsubsection*{$(M,M')=(3,3)$}
\begin{align}
\widehat{\cC}^{(3, 3)}_{N,p}(\tau_2) &=\left[ 1+\frac{2p}{3}\right] \frac{18 N \zeta(3)}{\tau_2 \pi}-\left[1+ { \left(5N^2+67\right) p \over \left(N^2{+}5\right) \left(N^2{+}7\right)  } \left( \frac{ N^2+5 }{6 }+\frac{ p}{3  } \right) \right] \frac{90 N^2 \zeta(5)}{\tau_2^{2} \pi^{2}} \\
&+\left[ 1+ {\left(N^2+23\right) p \over  \left(N^2 {+} 5\right)  \left(N^2 {+} 7\right) \left(N^2 {+} 9\right) } \left( \frac{ 3 N^4+33 N^2+94 }{3  }+5\left(N^2+5\right) p+\frac{20  p^2}{3 } \right) \right]\frac{735 N^3\zeta(7)}{2 \tau_2^{3} \pi^{3}} +\ldots \, . \nonumber
\end{align}
\subsubsection*{$(M,M')=(4,4)$}
\ie
\widehat{\cC}^{(4, 4)}_{N,p}(\tau_2) &=\left[ 1+\frac{p}{2}\right] \frac{24 N \zeta(3)}{\tau_2 \pi}-\left[ 1+\frac{\left(5 N^6+96 N^4-5 N^2+144\right) p}{ \left(N^2{+}7\right)
   \left(N^2{+}9\right) \left(2 N^4 {+} 3\right)}  \left( {  N^2+7  \over 4}+ {p \over 2}  \right) \right] \frac{60 \left(2N^{4}+3\right) \zeta(5)}{\left(N^2+1\right) \tau_2^{2} \pi^{2}} \cr
&+\left[ 1+ \frac{ \left(7 N^6+240 N^4-367 N^2+1080\right) p}{\left(N^2 {+} 7\right) \left(N^2 {+} 9\right) \left(N^2 {+} 11\right) \left(2 N^4 {-} 3 N^2 {+} 9\right)}  \left( \frac{3 N^4+45 N^2+172 }{15 }+\left(N^2+7\right)p +\frac{4 p^2}{3 } \right) \right]  \cr   
& \frac{525 N \left(2N^4-3N^2+9\right) \zeta(7)}{2\left(N^2+1\right) \tau_2^{3} \pi^{3}} +\ldots \, . 
\fe
\subsubsection*{$(M,M')=(5,5)$}
\ie
\widehat{\cC}^{(5, 5)}_{N,p}(\tau_2) &=\left[ 1+\frac{2p}{5}\right] \frac{30 N \zeta(3)}{\tau_2 \pi} -\left[ 1+ \frac{\left(N^6+28 N^4+43 N^2+144\right) p}{\left(N^2 {+} 9\right) \left(N^2 {+} 11\right)
   \left(N^4 {+} 2 N^2 {+} 6\right)}  \left( \frac{N^2+9 }{2 }+p \right) \right] \frac{150 \left(N^{4} {+} 2N^2{+}6\right) \zeta(5)}{\left(N^2+5\right) \tau_2^{2} \pi^{2}}   \cr
&+\left [ 1+ \frac{ \left(7 N^6+328 N^4-515 N^2+5220\right) p}{ \left(N^2 {+} 9\right)
   \left(N^2 {+} 11\right) \left(N^2 {+} 13\right) \left(5 N^4 {-} 8 N^2 {+} 87\right)}   \left( \frac{2 \left(3 N^4+57 N^2+274\right) }{15} +2 \left(N^2+9\right) p +\frac{8   p^2}{3} \right) \right]
   \cr
&\frac{525 N \left(5N^4-8N^2+87\right) \zeta(7)}{4\left(N^2+5\right) \tau_2^{3} \pi^{3}}+\ldots \, .
\fe

It is easy to check that the above results are solutions to the Laplace-difference equation. When $p=0$, the above expressions reproduce the results for the integrated correlators involving the single-particle operators that were considered in \cite{Paul:2022piq}, which also serve as the initial conditions for the Laplace-difference equation. These particular examples clearly show some intriguing structures of the perturbative expansion. For example, we observe that the leading term in all the examples we studied  takes the simple form of ${6 N } \left({2p} + M \right)\zeta(3)/(\tau_2 \, \pi) $. 

Finally, to obtain $\mathcal{\cC}_{N,p}^{(M,M'|i,i')}$, we will also need the expression for $\widetilde{R}_{N,p}^{(M, M'|i,i')}$ that appeared in the definition \eqref{eq:newC}.  Recall for the identical operators, namely the cases with $M=M'$ and $i=i'$, $\widetilde{R}_{N,p}^{(M, M|i,i)}={R}_{N,p}^{(M, M|i,i)}$, which is determined in terms of the two-point function as given in \eqref{eq:RMMp}. Because two-point functions are protected by supersymmetry, it is straightforward to compute them simply by Wick contractions. For example, for a particular set of operators $\mathcal{O}^{(i)}_{p|M} = (T_2)^p \, \mathcal{S}_M$ with $\mathcal{S}_M$ being the single-particle operators \cite{Aprile:2020uxk} that are defined in \eqref{eq:OmTmulti}, we find 
       \ie \label{eq:Rgen}
R_{N,p}^{(\mathcal{S}_M)}=R_{N,0}^{(\mathcal{S}_M)} {M^{\epsilon}\times 2^{2 p} \, p! \over (2 p+M)^2 } \left(\frac{N^2+2M-1}{2}\right){}_{p}\, , 
   \fe
where $\epsilon=2$ when $M>0$ and $\epsilon=1$ if $M \rightarrow 0$, and $R_{N;0}^{(\mathcal{S}_M)}$ is given by \cite{Aprile:2020uxk}
\ie
R_{N,0}^{(\mathcal{S}_M)} =  (M-1) \left[ \frac{1}{(N-M+1)_{M-1}}-\frac{1}{(N+1
   )_{M-1}} \right]^{-1}\, .
   \fe
The cases with $M=3, 4, 5$ are relevant to the integrated correlators we consider here. 

\subsection{Integrated correlators with $M \neq M'$}

In this subsection we consider the examples of integrated correlators with $M \neq M'$, for $M, M' \leq 5$.   First, we note all the integrated correlators $\widehat{\cC}^{(M, M'|i, i')}_{N,p}$ with $M+M'$ being an odd number always vanish. Therefore we only need to consider the cases with $(M,M')=(4,0)$ and $(M,M')=(5,3)$, which we will list below: 
\subsubsection*{$(M,M')=(4,0)$}
\ie
\!\!\! \!\!\! \! \! \! \!  \widehat{\cC}^{(4, 0)}_{N,p}(\tau_2)&=-\left[1+\frac{4 p \left( (N^2+4) +p\right)}{\left(N^2+3\right) \left(N^2+5\right)} \right] \!\!  \frac{120 N \zeta(5)}{\tau_2^{2} \pi^{2}}+\left[1+\frac{4 p \left( (2 N^4  + 19 N^2 +43) + (5 N^2 + 23 ) p + 4 p^2 \right)}{\left(N^2+3\right)
   \left(N^2+5\right) \left(N^2+7\right)}\right] \! \! \frac{1470 N^{2} \zeta(7)}{\tau_2^{3} \pi^{3}} \cr 
&-\left[1+\frac{4 p \left(\left(N^2+7\right) \left(10 N^4+95 N^2+213\right)+\left(3 N^2+13\right) \left(15 N^2+91\right) p+84 \left(N^2+5\right) p^2+56 p^3 \right)}{3 \left(N^2+3\right) \left(N^2+5\right) \left(N^2+7\right)
   \left(N^2+9\right)}\right] \cr
&\frac{2835 N^{}\left(4 N^{2}+1\right) \zeta(9)}{\tau_2^{4} \pi^{4}}+\ldots \, .
\fe
\subsubsection*{$(M,M')=(5,3)$}
\ie
\!\!\! \!\!\!\! \!\! \!\!  \widehat{\cC}^{(5, 3)}_{N,p} (\tau_2)&=-\left[1+\frac{2 p \left( (3 N^2+23)+4 p\right)}{\left(N^2+7\right) \left(N^2+9\right)}\right]\frac{180 N \zeta(5)}{\tau_2^{2} \pi^{2}}+\left[1+  \frac{4 p \left( \left(3 N^2 {+} 23\right)\left(N^2 {+} 9\right)  {+} 10 p \left(N^2 {+} 8\right) {+} 10p^2\right) }{\left(N^2+7\right) \left(N^2+9\right)
   \left(N^2+11\right)}   \right]\frac{1890 N^{2} \zeta(7)}{\tau_2^{3} \pi^{3}} \cr
&-\left[1+\frac{4 p \left(\left(5 N^6 {+} 135 N^4 {+} 1215 N^2 {+} 3629\right)+2 \left(15 N^4 {+} 255 N^2 {+} 1088\right) p+14 \left(5 N^2 {+} 41\right) p^2+56
   p^3 \right)}{\left(N^2+7\right) \left(N^2+9\right) \left(N^2+11\right)
   \left(N^2+13\right)}\right] \cr
&\frac{8605 N^{}\left(3 N^{2}+1\right) \zeta(9)}{2 \tau_2^{4} \pi^{4}} +\ldots \, .
\fe
Again, we have omitted the degeneracy index $i$. It is also straightforward to verify that these are solutions to the Laplace-difference equation \eqref{eq:gen-LDu}, and in appendix \ref{app:Nrec} we also provide the initial conditions for the recursion relation, namely $\widehat{\cC}^{(4, 0)}_{N,0}(\tau, \bar \tau)$ and  $\widehat{\cC}^{(5, 3)}_{N,0}(\tau, \bar \tau)$ for any $N$ and $\tau$. The perturbative expansion of the integrated correlators obeys interesting structures that are similar to those with $M=M'$.  However, we also note that interestingly the integrated correlators with operators that are in different (sub)towers in general start at higher loops in the perturbative expansion. For the examples we considered here, they both begin at two loops.  In a related context, a similar phenomenon was observed for the $U(1)$-violating correlators in $\mathcal{N}=4$ SYM studied in \cite{Green:2020eyj, Dorigoni:2021rdo}. 

Again, to obtain $\mathcal{\cC}_{N,p}^{(M,M'|i,i')}$, we now compute the normalisation factor $\widetilde{R}_{N;p}^{(M, M'|i,i')}$. For the integrated correlators with operators in different (sub)towers, we use the expression given in \eqref{eq:Rt1}. For  $M=4, M'=0$, we find
\ie  
\widetilde{R}^{(4,0)}_{N,p} 
& =  {  2^{2p-2} \,  (p+2)! \over  (p+2)^{2}  }  \left(\frac{N^2+3}{2}\right)_p  \,  \frac{\left(N^2-9\right) \left(N^2-4\right)
   \left(N^2-1\right)}{ \left(N^2+1\right)}\, , 
\fe
and for $M=5, M'=3$, we have
\ie  
\widetilde{R}^{(5,3)}_{N,p} 
= \frac{2^{2p-1} (p+1)!}{(2p+5)^2}
   \left(\frac{N^2+7}{2}\right)_p \,  \frac{5\left(N^2-16\right) \left(N^2-9\right)
   \left(N^2-4\right) \left(N^2-1\right)}{N
   \left(N^2+5\right)}  \, .
\fe

\subsection{$SL(2, \mathbb{Z})$ completion and lattice-sum representation}

In the previous subsections we presented examples of the perturbative expansions (i.e. zero instanton sector) of integrated correlators. The complete results of integrated correlators are $SL(2, \mathbb{Z})$ modular invariant functions of $(\tau, \bar{\tau})$ because of Montonen-Olive duality of $\mathcal{N}=4$ SYM with $SU(N)$ gauge group \cite{Montonen:1977sn}.  Therefore, the perturbative results should be completed into some modular functions, which in general is not unique at all. However,  as already noted in the case of $\langle \mathcal{O}_2 \mathcal{O}_2 \mathcal{O}_2 \mathcal{O}_2 \rangle$  in \cite{Dorigoni:2021guq}, by assuming the so-called lattice-sum representation of the integrated correlator, which has been proved recently in \cite{Dorigoni:2022cua}, one can show that the integrated correlator is completely determined in terms of the zero-instanton perturbative results \cite{Dorigoni:2021guq, Dorigoni:2022zcr}\footnote{See \cite{Dorigoni:2022zcr} (and also \cite{ Alday:2021vfb} for the earlier work) on understanding the modular properties of the integrated correlator associated with $\langle \mathcal{O}_2 \mathcal{O}_2 \mathcal{O}_2 \mathcal{O}_2 \rangle$ in $\mathcal{N}=4$ SYM with other classical gauge groups, and the manifestations of Goddard, Nuyts and Olive duality \cite{Goddard:1976qe}. }.  Essentially the same property was assumed in \cite{Aprile:2020uxk} (using the language of $SL(2, \mathbb{Z})$ spectral decomposition following \cite{Collier:2022emf}) for the more general integrated correlators we are considering here, which was verified by explicit perturbative and non-perturbative results \cite{Aprile:2020uxk}.

 Here we will briefly discuss the $SL(2, \mathbb{Z})$ completion of the integrated correlators and  their lattice representation. In a forthcoming paper \cite{largep}, we will study in detail the lattice representation  of the integrated correlators and its implications, such as  the modular properties of integrated correlators, and their behaviour in the large-charge limit.  Following \cite{Dorigoni:2021guq}, we propose that $\widehat{\mathcal{\cC}}_{N,p}^{(M,M'|i,i')}(\tau, \bar \tau)$ can be expressed as
\ie \label{eq:lattice}
\widehat{\mathcal{\cC}}_{N,p}^{(M,M'|i,i')} (\tau, \bar \tau)= \sum_{(m,n)\in \mathbb{Z}^2} \int_0^{\infty} e^{-t Y_{m,n}(\tau, \bar \tau) } \widehat{B}_{N,p}^{(M,M'|i,i')} (t) \, dt \, ,
\fe
with $Y_{m,n}(\tau, \bar \tau) := \pi \frac{|m+n\tau|^2}{\tau_2}$. The above expression is manifestly $SL(2, \mathbb{Z})$ invariant, and all the non-trivial information of the integrated correlators is encoded in the single rational function of $t$, $\widehat{B}_{N,p}^{(M,M'|i,i')} (t)$. The function $\widehat{B}_{N,p}^{(M,M'|i,i')} (t)$ obeys several important properties, for example, 
\ie
\widehat{B}_{N,p}^{(M,M'|i,i')} (t) = {1 \over t} \widehat{B}_{N,p}^{(M,M'|i,i')} (1/t) \, , \qquad  \int_0^{\infty} {dt \over \sqrt{t}} \widehat{B}_{N,p}^{(M,M'|i,i')} (t) =0 \, .
\fe
What is relevant to the current discussion is that, using \eqref{eq:lattice}, one can show that $\widehat{B}_{N,p}^{(M,M'|i,i')} (t)$ is uniquely determined by the perturbative results of $\widehat{\mathcal{\cC}}_{N,p}^{(M,M'|i,i')} (\tau, \bar \tau)$ \cite{Dorigoni:2021guq, Dorigoni:2022zcr}.  Furthermore, as the Laplace-difference equation \eqref{eq:gen-LDu} is also in an $SL(2, \mathbb{Z})$-invariant form, one may solve the equation and obtain modular functions for the integrated correlators, providing the $SL(2, \mathbb{Z})$-invariant initial data. Following the ideas of \cite{Dorigoni:2022cua}, one can in fact further introduce $SL(2, \mathbb{Z})$-invariant generating functions, that sum over all the charge-$p$ dependence of the integrated correlators.  The concept of the generating functions for the integrated correlators and their explicit expressions and applications will be studied in detail in \cite{largep}.

 \section{Conclusion and outlook}
 \label{sec:conclusion}

 In this paper we studied integrated correlators in $SU(N)$ $\mathcal{N}=4$ SYM associated with four-point correlation functions of the form $\langle \mathcal{O}_2 \mathcal{O}_2 \mathcal{O}^{(i)}_p \mathcal{O}^{(j)}_p \rangle$, for the half-BPS operators $\mathcal{O}^{(i)}_p$ with charge (or dimension) $p$.  To systematise  the supersymmetric localisation computation, we reorganised the operators into different towers as well as subtowers of the form $ \mathcal{O}^{(i)}_{p|M} = (\mathcal{O}_2)^p  \mathcal{O}^{(i)}_{0|M}$, with index $M$ denoting the tower and $i$ the subtower. Crucially, operators in different towers (and subtowers) are orthogonal to each other, which  greatly simplifies the Gram-Schmidt procedure for the localisation computation.   We proved that, remarkably, all the integrated correlators for any $N$ satisfy a universal Laplace-difference equation that relates integrated correlators of different charges $p$. It is vitally important to normalise the integrated correlators appropriately for them to obey the Laplace-difference equation. This is especially important when the integrated correlators involve two different operators, as a naive choice of the normalisation factor would either lead to $0/0$, or some very complicated Laplace-difference equation that would obscure the underlying simplicity of the integrated correlators. 
 
An analogous Laplace-difference equation, i.e. \eqref{eq:N-LG}, was found in \cite{Dorigoni:2021guq} for the integrated correlator associated with $\langle \mathcal{O}_2 \mathcal{O}_2 \mathcal{O}_2 \mathcal{O}_2 \rangle$, which relates the integrated correlator of different gauge groups.  Even though the equations may look similar, they seem to arise from different origins. The proof of the Laplace-difference equation of  \cite{Dorigoni:2021guq}, as was done recently in \cite{Dorigoni:2022cua}, crucially relies on the explicit form of the partition function of $\mathcal{N}=2^*$ SYM that enters in the localisation computation. As we showed, in proving the Laplace-difference equation of this paper which relates integrated correlators with different charges, the precise form of the partition function is not important. Our Laplace-difference equation in this sense is more similar to the Toda equations for the extremal correlators in $\mathcal{N}=2$ supersymmetric theories \cite{Gerchkovitz:2016gxx, Baggio:2014ioa, Baggio:2014sna, Baggio:2015vxa}. In both cases, the recursion relations relate correlators of operators with different charges, and the validity of recursion relations does not rely on the precise form of the partition functions that enter in the localisation computation. 

The Laplace-difference equation provides a powerful recursion relation that paves the way for determining  the integrated correlators in terms of initial data. To completely determine all the integrated correlators, one would also need to compute all initial conditions, i.e. $\widehat{\cC}^{(M,M'|i,i')}_{N,p}(\tau, \bar \tau)$ with $p=0$. We provided some examples of these initial conditions in appendix \ref{app:Nrec}. They seem to also obey some interesting recursion relations relating integrated correlators with different $N$ \cite{Paul:2022piq}. It will be very interesting to study these initial conditions systematically and understand better the recursion relations that are satisfied by them.

The unintegrated  correlators $\langle \mathcal{O}_2 \mathcal{O}_2 \mathcal{O}^{(i)}_{p} \mathcal{O}^{(j)}_{p}  \rangle$ have been studied extensively in various limits in the literature.  In the perturbative region, the correlators are known up to two loops for general $p$ \cite{DAlessandro:2005fnh} (and three loops in the planar limit \cite{Chicherin:2015edu}), and up to three loops for $p=2$ \cite{Eden:2012tu}. The integrands however are known to higher loops \cite{Fleury:2019ydf, Chicherin:2018avq, Bourjaily:2016evz}. In the strong coupling limit, they have been computed using bootstrap approaches and holographically in the planar limit \cite{Rastelli:2016nze,Rastelli:2017udc, Arutyunov:2018tvn,Caron-Huot:2018kta, Alday:2018pdi,Drummond:2019odu,Drummond:2020dwr,Abl:2020dbx,Aprile:2020mus,Alday:2022uxp}\footnote{See also \cite{Goncalves:2023oyx} for the recent results of the extension to the five-point correlator $\langle \mathcal{O}_2 \mathcal{O}_2 \mathcal{O}_2 \mathcal{O}^{(i)}_{p} \mathcal{O}^{(j)}_{p}  \rangle$  in the supergravity limit.}  and beyond \cite{Alday:2017xua,Aprile:2017bgs,Alday:2017vkk,Aprile:2017qoy,Aprile:2019rep,Alday:2019nin,Drummond:2020uni,Huang:2021xws,Drummond:2022dxw}. Our results provide additional important information about these correlators that are exact with finite $\tau$ and make the $SL(2,\mathbb{Z})$ symmetry manifest. It is  feasible that one could reconstruct unintegrated correlators by writing down an appropriate ansatz, probably in the large-$p$ limit, and to use our results as constraints to fix the ansatz. It would also be interesting to exploit the higher-loop integrands to study the relations between the perturbative parts of our results and these Feynman integrals using the connections between integrated correlators and Feynman integral periods observed in \cite{Wen:2022oky}. It would be particularly interesting to understand from the standard Feynman integral point of view the absence of the lower-loop terms in the perturbative expansion of the integrated correlators involving operators in different (sub)towers. 

Another natural question that arises from our results is to solve the Laplace-difference equation, and to obtain explicit expressions for the integrated correlators and to understand the dependence on the charge $p$. Following the ideas of \cite{Dorigoni:2022cua},  this can be conveniently done by introducing the generating functions of the integrated correlators, which sum over the charge-$p$ dependence. In a forthcoming paper \cite{largep}, we will see that the Laplace-difference equation becomes a differential equation for the generating functions, and we will solve the equation explicitly and derive the generating functions of the integrated correlators. We will be particularly interested in the large-$p$ behaviour of the integrated correlators.  From the generating functions, we will find that in the large-$p$ expansion, the integrated correlators take a universal form, which resembles the large-$N$ expansion of the integrated correlators \cite{Dorigoni:2022cua}\footnote{See also \cite{Hatsuda:2022enx} for the related work on the large-$N$ expansion of the integrated correlator $\cC_{N,1}^{(0,0)}(\tau, \bar \tau)$ in the 't Hooft genus expansion and the associated large-$N$ non-perturbative corrections using the resurgence analysis.} with $p$ and $N$ exchanged, albeit with some interesting and subtle differences.

\section*{Acknowledgements}

The authors would like to thank Daniele Dorigoni, Michael Green and Rodolfo Russo for insightful discussions. 
CW is supported by Royal Society University Research Fellowships No.~UF160350 and URF$\backslash$R$\backslash$221015. AB is supported by a Royal Society funding No.~RF$\backslash$ERE$\backslash$210067.  

\appendix

\section{Initial conditions for  Laplace-difference equation}\label{app:Nrec}

In this appendix, we will provide exact results for the integrated correlators $\widehat{\cC}^{(M,M'|i,i')}_{N,p}$ with $p=0$ for $M, M' \leq 5$, namely the initial conditions for the Laplace-difference equation for the examples  that we considered in section \ref{sec:local}. It was found in \cite{Paul:2022piq} that the integrated correlators  associated with $\langle \mathcal{O}_2 \mathcal{O}_2 \mathcal{O}^{(i)}_p \mathcal{O}^{(j)}_p \rangle$ for $p \leq 5$ obey some recursion relations that relate these integrated correlators with different $N$.  As we discussed in section \ref{sec:re-org}, the towers of operators $\mathcal{O}^{(i)}_{0|M}$ are particular linear combinations of  $\mathcal{O}^{(i)}_p$ or $T_{p_1, p_2, \ldots, p_n}$. For the relevance of this discussion, the examples given in \eqref{eq:exOm} are quoted below,
\ie \label{eq:exOm3}
\mathcal{O}_{0|0} =\mathbb{I}\, , \qquad \mathcal{O}_{0|3} = T_3\, , \qquad \mathcal{O}_{0|4} = T_4- {2N^2-3\over N(N^2+1)} T_{2,2}\, ,
\qquad \mathcal{O}_{0|5} = T_5- {5(N^2-2)\over N(N^2+1)} T_{2,3} \, .
\fe
As we commented, they also coincide with the identical single-particle operators. The integrated correlators involving two identical single-particle operators with $M<6$ have been considered in \cite{Paul:2022piq}. In our notation, they are $\widehat{\cC}^{(M,M|i,i)}_{N,0}(\tau, \bar \tau)$, with  $M=3,4,5$. Therefore, here we will only consider the cases with $M\neq M'$. Since there is no degeneracy we will also drop the indices $i,i'$, and the only non-trivial cases are $\widehat{\cC}^{(4, 0)}_{N,0}(\tau, \bar \tau)$ and $\widehat{\cC}^{(5, 3)}_{N,0}(\tau, \bar \tau)$, which we will present below.

Using \eqref{eq:exOm3}, we have the following relation between four-point functions
\ie \label{eq:relation}
\langle \mathcal{O}_2\, \mathcal{O}_2 \,\mathcal{O}_{0|4}\, \mathcal{O}_{2|0} \rangle = \langle \mathcal{O}_2 \, \mathcal{O}_2 \, T_4 \, T_{2,2} \rangle - {2N^2-3\over N(N^2+1)} \langle \mathcal{O}_2 \, \mathcal{O}_2 \, T_{2,2} \, T_{2,2} \rangle \, , 
\fe
which implies the same relation for the corresponding integrated correlators. It was  found in \cite{Paul:2022piq} (see section 3 of the reference) that the integrated correlators associated with $\langle \mathcal{O}_2 \, \mathcal{O}_2 \, T_4 \, T_{2,2} \rangle$ and $\langle \mathcal{O}_2 \, \mathcal{O}_2 \, T_{2,2} \, T_{2,2} \rangle$ obey some $N$-dependent recursion relations. Solving these recursion relations, using \eqref{eq:relation} and adapting to our normalisation, we obtain
the results for $\widehat{\cC}^{(4, 0)}_{N,0}(\tau, \bar \tau)$ in terms of only $\mathcal{\cC}^{(0, 0)}_{N,1}(\tau, \bar \tau)$. Once again, $\mathcal{\cC}^{(0, 0)}_{N,1}(\tau, \bar \tau)$ is known exactly \cite{Dorigoni:2021bvj, Dorigoni:2021guq}, which may be determined by the recursion relation \eqref{eq:N-LG}.  Explicitly, we find
\ie
\widehat{\cC}^{(4, 0)}_{N,0}(\tau, \bar \tau)=\frac{\left(N-3\right)_7}{32N(N^2+1)} \left[{F}_N^{(4)}(\tau, \bar \tau) -\frac{2N^2-3}{N(N^2+1)}\left(\Delta_{\tau}+2N^2 \right)\mathcal{\cC}^{(0, 0)}_{N,1}(\tau, \bar \tau) \right] \, ,
\fe
where $\left(N-3\right)_7$ is the Pochhammer symbol.
Using the Laplace-difference equation \eqref{eq:N-LG}, one could replace $\Delta_{\tau}\mathcal{\cC}_{N}(\tau, \bar \tau)$ by linear combinations of $\mathcal{\cC}^{(0, 0)}_{N-1,1}(\tau, \bar \tau)$, $\mathcal{\cC}^{(0, 0)}_{N,1}(\tau, \bar \tau)$ and $\mathcal{\cC}^{(0, 0)}_{N+1,1}(\tau, \bar \tau)$.  Furthermore, the function ${F}_N^{(4)}(\tau, \bar \tau)$ is also expressed in terms of $\mathcal{\cC}^{(0, 0)}_{N,1}(\tau, \bar \tau)$, 
\ie \label{eq:FN4}
{F}_N^{(4)}(\tau, \bar \tau)=\sum_{l=1}^{N-1} \frac{1}{2l(l+1)} &\left[l^2(l+1)^2 \left(\mathcal{\cC}^{(0, 0)}_{l+2,1}(\tau, \bar \tau)+\mathcal{\cC}^{(0, 0)}_{l-1,1}(\tau, \bar \tau)\right)-l (l-2) \left( l^2+4l -1\right) \mathcal{\cC}^{(0, 0)}_{l+1,1}(\tau, \bar \tau) \right.\cr 
&\left.- (l+1) (l+3) \left(l^2-2 l-4\right) \mathcal{\cC}^{(0, 0)}_{l,1}(\tau, \bar \tau)\right] \, .
\fe
Similarly for $\widehat{\cC}^{(5, 3)}_{N,0}(\tau, \bar \tau)$, we have
\ie
&\frac{40N^2 (N^2+5)}{\left(N-4\right)_9}\widehat{\cC}^{(5, 3)}_{N,0} (\tau, \bar \tau)=   {F}_N^{(5)}(\tau, \bar \tau)+\frac{6}{25N(N-1)}\left[3N(N-1)(N^2+17) {\cC}^{(3,3)}_{N-1,0}(\tau, \bar \tau) \right.\cr
&\left.-6N(N+1)(N^2+2N-21)\mathcal{\cC}^{(0, 0)}_{N-1,1}(\tau, \bar \tau)+2(N-1)(N-2)\left((N^3+5N^2-13N+25)\mathcal{\cC}^{(0, 0)}_{N,1}(\tau, \bar \tau) \right. \right. \cr
& \left. \left. +6N(N-1)\mathcal{\cC}^{(0, 0)}_{N+1,1}(\tau, \bar \tau)\right)\right] \, ,
\fe
where ${\cC}^{(3,3)}_{N,0}(\tau, \bar \tau)$ is the integrated correlator associated with $\langle \mathcal{O}_2 \mathcal{O}_2 \mathcal{O}_3 \mathcal{O}_3  \rangle$, which is given by
\ie \label{eq:solp3}
{\cC}^{(3,3)}_{N,0}(\tau, \bar \tau) = 2 \sum_{l=1}^{N-1}  \left( \cC^{(0, 0)}_{l,1} (\tau, \bar \tau)+ \cC^{(0, 0)}_{l+1,1}(\tau, \bar \tau)\right) - {4\over N} \cC^{(0, 0)}_{N,1}(\tau, \bar \tau) \, ,
\fe
and ${F}_N^{(5)}(\tau, \bar \tau)$ taking the following form,  
\ie
\!\!\!\!\!\!\! {F}_N^{(5)}(\tau, \bar \tau)=\sum_{l=1}^{N-1} &\frac{6}{5l^2 (l+1)^2}\! \left[4l(l+2)(l+1)^2 {F}_N^{(4)}(\tau, \bar \tau) + l(l^2-1)\left((9l-6) \cC^{(3,3)}_{N,0}(\tau, \bar \tau)+2l^2(l+1) \mathcal{\cC}^{(0, 0)}_{N+2,1}(\tau, \bar \tau) \right)\right.\cr 
&\left.-2l^2(l-1)(3l^2-4l+3)\mathcal{\cC}^{(0, 0)}_{N+1,1}(\tau, \bar \tau) -2(l+1)(l+2)(l^4+l^3-11l^2+11l-8)\mathcal{\cC}^{(0, 0)}_{N,1}(\tau, \bar \tau)\right] \, ,
\fe
with ${F}_N^{(4)}(\tau, \bar \tau)$ and ${\cC}^{(3,3)}_{N,0}(\tau, \bar \tau)$ given in \eqref{eq:FN4} and \eqref{eq:solp3}, respectively. 

	\bibliographystyle{ssg}
	\bibliography{genfun2}

\begingroup\raggedright\begin{thebibliography}{10}

\bibitem{Binder:2019jwn}
D.~J. Binder, S.~M. Chester, S.~S. Pufu, and Y.~Wang, ``{$ \mathcal{N} $ = 4
  Super-Yang-Mills correlators at strong coupling from string theory and
  localization},'' {\em JHEP} {\bf 12} (2019) 119,
  \href{https://arxiv.org/abs/1902.06263}{{\tt 1902.06263}}.

\bibitem{Chester:2020dja}
S.~M. Chester and S.~S. Pufu, ``{Far beyond the planar limit in
  strongly-coupled $ \mathcal{N} $ = 4 SYM},'' {\em JHEP} {\bf 01} (2021) 103,
  \href{https://arxiv.org/abs/2003.08412}{{\tt 2003.08412}}.

\bibitem{Chester:2019pvm}
S.~M. Chester, ``{Genus-2 holographic correlator on AdS$_{5} \times$ S$^{5}$
  from localization},'' {\em JHEP} {\bf 04} (2020) 193,
  \href{https://arxiv.org/abs/1908.05247}{{\tt 1908.05247}}.

\bibitem{Chester:2019jas}
S.~M. Chester, M.~B. Green, S.~S. Pufu, Y.~Wang, and C.~Wen, ``{Modular
  invariance in superstring theory from $ \mathcal{N} $ = 4
  super-Yang-Mills},'' {\em JHEP} {\bf 11} (2020) 016,
  \href{https://arxiv.org/abs/1912.13365}{{\tt 1912.13365}}.

\bibitem{Chester:2020vyz}
S.~M. Chester, M.~B. Green, S.~S. Pufu, Y.~Wang, and C.~Wen, ``{New modular
  invariants in $ \mathcal{N} $ = 4 Super-Yang-Mills theory},'' {\em JHEP} {\bf
  04} (2021) 212, \href{https://arxiv.org/abs/2008.02713}{{\tt 2008.02713}}.

\bibitem{Pestun:2007rz}
V.~Pestun, ``{Localization of gauge theory on a four-sphere and supersymmetric
  Wilson loops},'' {\em Commun. Math. Phys.} {\bf 313} (2012) 71--129,
  \href{https://arxiv.org/abs/0712.2824}{{\tt 0712.2824}}.

\bibitem{Nekrasov:2002qd}
N.~A. Nekrasov, ``{Seiberg-Witten prepotential from instanton counting},'' {\em
  Adv. Theor. Math. Phys.} {\bf 7} (2003), no.~5 831--864,
  \href{https://arxiv.org/abs/hep-th/0206161}{{\tt hep-th/0206161}}.

\bibitem{Dorigoni:2021bvj}
D.~Dorigoni, M.~B. Green, and C.~Wen, ``{Novel Representation of an Integrated
  Correlator in $\mathcal N$ = 4 Supersymmetric Yang-Mills Theory},'' {\em
  Phys. Rev. Lett.} {\bf 126} (2021), no.~16 161601,
  \href{https://arxiv.org/abs/2102.08305}{{\tt 2102.08305}}.

\bibitem{Dorigoni:2021guq}
D.~Dorigoni, M.~B. Green, and C.~Wen, ``{Exact properties of an integrated
  correlator in $ \mathcal{N} $ = 4 SU(N) SYM},'' {\em JHEP} {\bf 05} (2021)
  089, \href{https://arxiv.org/abs/2102.09537}{{\tt 2102.09537}}.

\bibitem{Dorigoni:2022zcr}
D.~Dorigoni, M.~B. Green, and C.~Wen, ``{Exact results for duality-covariant
  integrated correlators in $\mathcal{N}=4$ SYM with general classical gauge
  groups},'' {\em SciPost Phys.} {\bf 13} (2, 2022) 092,
  \href{https://arxiv.org/abs/2202.05784}{{\tt 2202.05784}}.

\bibitem{Dorigoni:2022iem}
D.~Dorigoni, M.~B. Green, and C.~Wen, ``{The SAGEX Review on Scattering
  Amplitudes, Chapter 10: Modular covariance of type IIB string amplitudes and
  their $\mathcal{N}=4$ supersymmetric Yang-Mills duals},''
  \href{https://arxiv.org/abs/2203.13021}{{\tt 2203.13021}}.

\bibitem{Paul:2022piq}
H.~Paul, E.~Perlmutter, and H.~Raj, ``{Integrated Correlators in
  $\mathcal{N}=4$ SYM via $SL(2,\mathbb{Z})$ Spectral Theory},''
  \href{https://arxiv.org/abs/2209.06639}{{\tt 2209.06639}}.

\bibitem{Rayson:2008uje}
C.~Rayson, {\em {Some aspects of conformal ${\cal N}=4$ SYM four point
  function}}.
\newblock PhD thesis, Cambridge U., 2008.
\newblock \href{https://arxiv.org/abs/1706.04450}{{\tt 1706.04450}}.

\bibitem{Aprile:2020uxk}
F.~Aprile, J.~M. Drummond, P.~Heslop, H.~Paul, F.~Sanfilippo, M.~Santagata, and
  A.~Stewart, ``{Single particle operators and their correlators in free $
  \mathcal{N} $ = 4 SYM},'' {\em JHEP} {\bf 11} (2020) 072,
  \href{https://arxiv.org/abs/2007.09395}{{\tt 2007.09395}}.

\bibitem{DHoker:2000xhf}
E.~D'Hoker, J.~Erdmenger, D.~Z. Freedman, and M.~Perez-Victoria, ``{Near
  extremal correlators and vanishing supergravity couplings in AdS / CFT},''
  {\em Nucl. Phys. B} {\bf 589} (2000) 3--37,
  \href{https://arxiv.org/abs/hep-th/0003218}{{\tt hep-th/0003218}}.

\bibitem{Fiol:2023cml}
B.~Fiol and Z.~Kong, ``{The planar limit of integrated 4-point functions},''
  \href{https://arxiv.org/abs/2303.09572}{{\tt 2303.09572}}.

\bibitem{Montonen:1977sn}
C.~Montonen and D.~I. Olive, ``{Magnetic Monopoles as Gauge Particles?},'' {\em
  Phys. Lett. B} {\bf 72} (1977) 117--120.

\bibitem{Gerchkovitz:2016gxx}
E.~Gerchkovitz, J.~Gomis, N.~Ishtiaque, A.~Karasik, Z.~Komargodski, and S.~S.
  Pufu, ``{Correlation Functions of Coulomb Branch Operators},'' {\em JHEP}
  {\bf 01} (2017) 103, \href{https://arxiv.org/abs/1602.05971}{{\tt
  1602.05971}}.

\bibitem{Erictoapp}
H.~Paul, E.~Perlmutter, and H.~Raj, ``{Exact Large Charge in $\mathcal{N}=4$
  SYM and Semiclassical String Theory},''
  \href{https://arxiv.org/abs/2303.13207}{{\tt 2303.13207}}.

\bibitem{Baggio:2012rr}
M.~Baggio, J.~de~Boer, and K.~Papadodimas, ``{A non-renormalization theorem for
  chiral primary 3-point functions},'' {\em JHEP} {\bf 07} (2012) 137,
  \href{https://arxiv.org/abs/1203.1036}{{\tt 1203.1036}}.

\bibitem{Eden:2000bk}
B.~Eden, A.~C. Petkou, C.~Schubert, and E.~Sokatchev, ``{Partial
  nonrenormalization of the stress tensor four point function in N=4 SYM and
  AdS / CFT},'' {\em Nucl. Phys. B} {\bf 607} (2001) 191--212,
  \href{https://arxiv.org/abs/hep-th/0009106}{{\tt hep-th/0009106}}.

\bibitem{Nirschl:2004pa}
M.~Nirschl and H.~Osborn, ``{Superconformal Ward identities and their
  solution},'' {\em Nucl. Phys. B} {\bf 711} (2005) 409--479,
  \href{https://arxiv.org/abs/hep-th/0407060}{{\tt hep-th/0407060}}.

\bibitem{Fucito:2015ofa}
F.~Fucito, J.~F. Morales, and R.~Poghossian, ``{Wilson loops and chiral
  correlators on squashed spheres},'' {\em JHEP} {\bf 11} (2015) 064,
  \href{https://arxiv.org/abs/1507.05426}{{\tt 1507.05426}}.

\bibitem{Russo:2013kea}
J.~G. Russo and K.~Zarembo, ``{Massive N=2 Gauge Theories at Large N},'' {\em
  JHEP} {\bf 11} (2013) 130, \href{https://arxiv.org/abs/1309.1004}{{\tt
  1309.1004}}.

\bibitem{Green:2020eyj}
M.~B. Green and C.~Wen, ``{Maximal U(1)$_{Y}$-violating n-point correlators in
  $ \mathcal{N} $ = 4 super-Yang-Mills theory},'' {\em JHEP} {\bf 02} (2021)
  042, \href{https://arxiv.org/abs/2009.01211}{{\tt 2009.01211}}.

\bibitem{Dorigoni:2021rdo}
D.~Dorigoni, M.~B. Green, and C.~Wen, ``{Exact expressions for $n$-point
  maximal $U(1)_Y$-violating integrated correlators in $SU(N)$ $\mathcal{N}=4$
  SYM},'' {\em JHEP} {\bf 11} (2021) 132,
  \href{https://arxiv.org/abs/2109.08086}{{\tt 2109.08086}}.

\bibitem{Dorigoni:2022cua}
D.~Dorigoni, M.~B. Green, C.~Wen, and H.~Xie, ``{Modular-invariant large-$N$
  completion of an integrated correlator in $\mathcal{N}=4$ supersymmetric
  Yang-Mills theory},'' \href{https://arxiv.org/abs/2210.14038}{{\tt
  2210.14038}}.

\bibitem{Alday:2021vfb}
L.~F. Alday, S.~M. Chester, and T.~Hansen, ``{Modular invariant holographic
  correlators for $ \mathcal{N} $ = 4 SYM with general gauge group},'' {\em
  JHEP} {\bf 12} (2021) 159, \href{https://arxiv.org/abs/2110.13106}{{\tt
  2110.13106}}.

\bibitem{Goddard:1976qe}
P.~Goddard, J.~Nuyts, and D.~I. Olive, ``{Gauge Theories and Magnetic
  Charge},'' {\em Nucl. Phys. B} {\bf 125} (1977) 1--28.

\bibitem{Collier:2022emf}
S.~Collier and E.~Perlmutter, ``{Harnessing S-duality in $ \mathcal{N} $ = 4
  SYM \& supergravity as SL(2, \ensuremath{\mathbb{Z}})-averaged strings},''
  {\em JHEP} {\bf 08} (2022) 195, \href{https://arxiv.org/abs/2201.05093}{{\tt
  2201.05093}}.

\bibitem{largep}
A.~Brown, C.~Wen, and H.~Xie, ``{Generating functions and large-charge
  expansion of integrated correlators in $\mathcal{N}=4$ supersymmetric
  Yang-Mills theory},'' \href{https://arxiv.org/abs/2303.17570}{{\tt
  2303.17570}}.

\bibitem{Baggio:2014ioa}
M.~Baggio, V.~Niarchos, and K.~Papadodimas, ``{tt$^{*}$ equations, localization
  and exact chiral rings in 4d $ \mathcal{N} $ =2 SCFTs},'' {\em JHEP} {\bf 02}
  (2015) 122, \href{https://arxiv.org/abs/1409.4212}{{\tt 1409.4212}}.

\bibitem{Baggio:2014sna}
M.~Baggio, V.~Niarchos, and K.~Papadodimas, ``{Exact correlation functions in
  $SU(2) \mathcal N=2$ superconformal QCD},'' {\em Phys. Rev. Lett.} {\bf 113}
  (2014), no.~25 251601, \href{https://arxiv.org/abs/1409.4217}{{\tt
  1409.4217}}.

\bibitem{Baggio:2015vxa}
M.~Baggio, V.~Niarchos, and K.~Papadodimas, ``{On exact correlation functions
  in SU(N) $ \mathcal{N}=2 $ superconformal QCD},'' {\em JHEP} {\bf 11} (2015)
  198, \href{https://arxiv.org/abs/1508.03077}{{\tt 1508.03077}}.

\bibitem{DAlessandro:2005fnh}
M.~D'Alessandro and L.~Genovese, ``{A Wide class of four point functions of BPS
  operators in N=4 SYM at order g**4},'' {\em Nucl. Phys. B} {\bf 732} (2006)
  64--88, \href{https://arxiv.org/abs/hep-th/0504061}{{\tt hep-th/0504061}}.

\bibitem{Chicherin:2015edu}
D.~Chicherin, J.~Drummond, P.~Heslop, and E.~Sokatchev, ``{All three-loop
  four-point correlators of half-BPS operators in planar $ \mathcal{N} $ = 4
  SYM},'' {\em JHEP} {\bf 08} (2016) 053,
  \href{https://arxiv.org/abs/1512.02926}{{\tt 1512.02926}}.

\bibitem{Eden:2012tu}
B.~Eden, P.~Heslop, G.~P. Korchemsky, and E.~Sokatchev, ``{Constructing the
  correlation function of four stress-tensor multiplets and the four-particle
  amplitude in N=4 SYM},'' {\em Nucl. Phys. B} {\bf 862} (2012) 450--503,
  \href{https://arxiv.org/abs/1201.5329}{{\tt 1201.5329}}.

\bibitem{Fleury:2019ydf}
T.~Fleury and R.~Pereira, ``{Non-planar data of $ \mathcal{N} $ = 4 SYM},''
  {\em JHEP} {\bf 03} (2020) 003, \href{https://arxiv.org/abs/1910.09428}{{\tt
  1910.09428}}.

\bibitem{Chicherin:2018avq}
D.~Chicherin, A.~Georgoudis, V.~Gon\c{c}alves, and R.~Pereira, ``{All five-loop
  planar four-point functions of half-BPS operators in $\mathcal N=4$ SYM},''
  {\em JHEP} {\bf 11} (2018) 069, \href{https://arxiv.org/abs/1809.00551}{{\tt
  1809.00551}}.

\bibitem{Bourjaily:2016evz}
J.~L. Bourjaily, P.~Heslop, and V.-V. Tran, ``{Amplitudes and Correlators to
  Ten Loops Using Simple, Graphical Bootstraps},'' {\em JHEP} {\bf 11} (2016)
  125, \href{https://arxiv.org/abs/1609.00007}{{\tt 1609.00007}}.

\bibitem{Rastelli:2016nze}
L.~Rastelli and X.~Zhou, ``{Mellin amplitudes for $AdS_5\times S^5$},'' {\em
  Phys. Rev. Lett.} {\bf 118} (2017), no.~9 091602,
  \href{https://arxiv.org/abs/1608.06624}{{\tt 1608.06624}}.

\bibitem{Rastelli:2017udc}
L.~Rastelli and X.~Zhou, ``{How to Succeed at Holographic Correlators Without
  Really Trying},'' {\em JHEP} {\bf 04} (2018) 014,
  \href{https://arxiv.org/abs/1710.05923}{{\tt 1710.05923}}.

\bibitem{Arutyunov:2018tvn}
G.~Arutyunov, R.~Klabbers, and S.~Savin, ``{Four-point functions of 1/2-BPS
  operators of any weights in the supergravity approximation},'' {\em JHEP}
  {\bf 09} (2018) 118, \href{https://arxiv.org/abs/1808.06788}{{\tt
  1808.06788}}.

\bibitem{Caron-Huot:2018kta}
S.~Caron-Huot and A.-K. Trinh, ``{All tree-level correlators in AdS$_{5} \times
  S^{5}$ supergravity: hidden ten-dimensional conformal symmetry},'' {\em JHEP}
  {\bf 01} (2019) 196, \href{https://arxiv.org/abs/1809.09173}{{\tt
  1809.09173}}.

\bibitem{Alday:2018pdi}
L.~F. Alday, A.~Bissi, and E.~Perlmutter, ``{Genus-One String Amplitudes from
  Conformal Field Theory},'' {\em JHEP} {\bf 06} (2019) 010,
  \href{https://arxiv.org/abs/1809.10670}{{\tt 1809.10670}}.

\bibitem{Drummond:2019odu}
J.~Drummond, D.~Nandan, H.~Paul, and K.~Rigatos, ``{String corrections to AdS
  amplitudes and the double-trace spectrum of $ \mathcal{N} $ = 4 SYM},'' {\em
  JHEP} {\bf 12} (2019) 173, \href{https://arxiv.org/abs/1907.00992}{{\tt
  1907.00992}}.

\bibitem{Drummond:2020dwr}
J.~M. Drummond, H.~Paul, and M.~Santagata, ``{Bootstrapping string theory on
  AdS$_5 \times S^5$},'' \href{https://arxiv.org/abs/2004.07282}{{\tt
  2004.07282}}.

\bibitem{Abl:2020dbx}
T.~Abl, P.~Heslop, and A.~E. Lipstein, ``{Towards the Virasoro-Shapiro
  amplitude in AdS$_{5} \times S^{5}$},'' {\em JHEP} {\bf 04} (2021) 237,
  \href{https://arxiv.org/abs/2012.12091}{{\tt 2012.12091}}.

\bibitem{Aprile:2020mus}
F.~Aprile, J.~M. Drummond, H.~Paul, and M.~Santagata, ``{The Virasoro-Shapiro
  amplitude in AdS$_{5} \times S^{5}$ and level splitting of 10d conformal
  symmetry},'' {\em JHEP} {\bf 11} (2021) 109,
  \href{https://arxiv.org/abs/2012.12092}{{\tt 2012.12092}}.

\bibitem{Alday:2022uxp}
L.~F. Alday, T.~Hansen, and J.~A. Silva, ``{AdS Virasoro-Shapiro from
  dispersive sum rules},'' {\em JHEP} {\bf 10} (2022) 036,
  \href{https://arxiv.org/abs/2204.07542}{{\tt 2204.07542}}.

\bibitem{Goncalves:2023oyx}
V.~Gon\c{c}alves, C.~Meneghelli, R.~Pereira, J.~Vilas~Boas, and X.~Zhou,
  ``{Kaluza-Klein Five-Point Functions from $\textrm{AdS}_5\times S_5$
  Supergravity},'' \href{https://arxiv.org/abs/2302.01896}{{\tt 2302.01896}}.

\bibitem{Alday:2017xua}
L.~F. Alday and A.~Bissi, ``{Loop Corrections to Supergravity on $AdS_5 \times
  S^5$},'' {\em Phys. Rev. Lett.} {\bf 119} (2017), no.~17 171601,
  \href{https://arxiv.org/abs/1706.02388}{{\tt 1706.02388}}.

\bibitem{Aprile:2017bgs}
F.~Aprile, J.~Drummond, P.~Heslop, and H.~Paul, ``{Quantum Gravity from
  Conformal Field Theory},'' {\em JHEP} {\bf 01} (2018) 035,
  \href{https://arxiv.org/abs/1706.02822}{{\tt 1706.02822}}.

\bibitem{Alday:2017vkk}
L.~F. Alday and S.~Caron-Huot, ``{Gravitational S-matrix from CFT dispersion
  relations},'' {\em JHEP} {\bf 12} (2018) 017,
  \href{https://arxiv.org/abs/1711.02031}{{\tt 1711.02031}}.

\bibitem{Aprile:2017qoy}
F.~Aprile, J.~Drummond, P.~Heslop, and H.~Paul, ``{Loop corrections for
  Kaluza-Klein AdS amplitudes},'' {\em JHEP} {\bf 05} (2018) 056,
  \href{https://arxiv.org/abs/1711.03903}{{\tt 1711.03903}}.

\bibitem{Aprile:2019rep}
F.~Aprile, J.~Drummond, P.~Heslop, and H.~Paul, ``{One-loop amplitudes in
  AdS$_{5} \times S^{5}$ supergravity from $ \mathcal{N} $ = 4 SYM at strong
  coupling},'' {\em JHEP} {\bf 03} (2020) 190,
  \href{https://arxiv.org/abs/1912.01047}{{\tt 1912.01047}}.

\bibitem{Alday:2019nin}
L.~F. Alday and X.~Zhou, ``{Simplicity of AdS Supergravity at One Loop},'' {\em
  JHEP} {\bf 09} (2020) 008, \href{https://arxiv.org/abs/1912.02663}{{\tt
  1912.02663}}.

\bibitem{Drummond:2020uni}
J.~Drummond, R.~Glew, and H.~Paul, ``{One-loop string corrections for AdS
  Kaluza-Klein amplitudes},'' \href{https://arxiv.org/abs/2008.01109}{{\tt
  2008.01109}}.

\bibitem{Huang:2021xws}
Z.~Huang and E.~Y. Yuan, ``{Graviton Scattering in
  $\mathrm{AdS}_5\times\mathrm{S}^5$ at Two Loops},''
  \href{https://arxiv.org/abs/2112.15174}{{\tt 2112.15174}}.

\bibitem{Drummond:2022dxw}
J.~M. Drummond and H.~Paul, ``{Two-loop supergravity on AdS$_{5} \times S^{5}$
  from CFT},'' {\em JHEP} {\bf 08} (2022) 275,
  \href{https://arxiv.org/abs/2204.01829}{{\tt 2204.01829}}.

\bibitem{Wen:2022oky}
C.~Wen and S.-Q. Zhang, ``{Integrated correlators in $ \mathcal{N} $ = 4 super
  Yang-Mills and periods},'' {\em JHEP} {\bf 05} (2022) 126,
  \href{https://arxiv.org/abs/2203.01890}{{\tt 2203.01890}}.

\bibitem{Hatsuda:2022enx}
Y.~Hatsuda and K.~Okuyama, ``{Large $N$ expansion of an integrated correlator
  in $\mathcal{N}=4$ SYM},'' \href{https://arxiv.org/abs/2208.01891}{{\tt
  2208.01891}}.

\end{thebibliography}\endgroup

\end{document}